\documentstyle[prl,aps]{revtex} \def\narrowtext{} \tighten\twocolumn

\begin{document}
\draft
\title{
\begin{minipage}[t]{7.0in}
\scriptsize
\begin{quote}
\leftline{{\it Phys. Rev.} {\bf B}, in press}
\raggedleft {\rm cond-mat/0406431}
\end{quote}
\end{minipage}
\medskip
\\Effects of the On-Site Coulomb Repulsion in Double Exchange Magnets}
\author{D. I. Golosov\thanks{Present address: Racah Institute of Physics, 
the Hebrew University, Jerusalem 91904, Israel. E-mail: golosov@phys.huji.ac.il}}
\address{Dept. of Condensed Matter Physics, Weizmann Institute of Science,
Rehovot 76100, Israel, and}
\address{Theoretical Physics, Oxford University, 1 Keble Rd., Oxford
OX1 3NP, United Kingdom} 
\address{%
\begin{minipage}[t]{6.0in}
\begin{abstract}
We investigate the zero-temperature phase diagram and spin-wave properties 
of a double exchange magnet with on-site Hubbard repulsion. 
It is shown that even within a simple Hartree -- Fock approach this interaction
(which is often omitted in theoretical treatments) leads to 
qualitatively important effects which are highly relevant in the
context of experimental data for the colossal magnetoresistance compounds.
These include the asymmetry of the doping dependence of spin stiffness, 
and the   zone-boundary
``softening'' of  spin wave dispersion. Effects of Hubbard repulsion on 
phase separation are analyzed as well.
We also show that in the ferromagnetic phase, an unusual 
temperature-dependent effective electron-electron interaction arises at
finite $T$. The mean-field scheme, however, does not yield the experimentally
observed density of states depletion near the Fermi level. 
We speculate that proper treatment of electron-electron
interactions may be necessary for understanding both this important feature 
and more generally the physics of colossal
magnetoresistance phenomenon. 
\typeout{polish abstract}
\end{abstract}
\end{minipage}}

\maketitle
\narrowtext
\newpage
\section{INTRODUCTION}
\label{sec:intro}

The phenomenon of colossal magnetoresistance (CMR) is known to occur in a
broad group of compounds, corresponding to different crystal structures, 
chemical compositions, and doping levels\cite{Tokurabook}. In addition to 
various heavily doped manganese oxides, the CMR effect is also observed in 
certain magnetic semiconductors and spinels\cite{Nagaevbook}; 
it is natural to expect that in all these cases, the physical 
origins of the CMR are similar. Thus, a proper minimal theoretical model
of a  CMR system should account for the important common features shared
by all these materials, while leaving out the peculiarities of crystal
environment and atomic structure of individual compounds. It is universally
recognised that one of these common features is the sizable ferromagnetic
Hund's rule coupling, $J_H$, between the spins of magnetic ions 
and those of conduction electrons, which gives rise to the double exchange
ferromagnetism\cite{Anderson} of the CMR compounds. The purpose of the 
present article is
to draw attention to the fact that another ubiquitous intra-atomic interaction,
namely the Coulomb (Hubbard) repulsion $U$, also affects magnetic, electronic, 
and transport properties of the system in a profound way, and may play a 
crucial role in the basic physics of the CMR. While some effects of this
interaction have been addressed in the past (see, {\it e.g.}, Refs. 
\cite{Mishra,Freeman,Marcelo,Kieinert}),  its potential importance is
not yet fully appreciated. We will argue that the on-site Coulomb repulsion
strongly affects the magnetic properties of the system; some generic 
experimental facts are recovered. We also suggest that the effects of
Hubbard repulsion merit further investigation beyond the mean-field
approach. 

We start with the standard double exchange Hamiltonian, supplemented with
a Hubbard repulsion term:

\begin{eqnarray}
{\cal H}=&-&\frac{t}{2} \sum_{\langle i,j \rangle,\alpha} 
\left(c^\dagger_{i \alpha}c_{j \alpha} +c^\dagger_{j \alpha}c_{i
\alpha}\right) - 
 \frac{J_H}{2S} \sum_{i, \alpha, \beta} \vec{S}_i
\vec{\sigma}^{\alpha \beta} c^\dagger_{i\alpha} c_{i\beta}
+ \nonumber \\
&+&\frac{J}{S^2}\sum_{\langle i, j \rangle} \vec{S}_i \vec{S}_j +
U \sum_i c^\dagger_{i \uparrow} c^\dagger_{i \downarrow}
c_{i \downarrow}  c_{i \uparrow}.
\label{eq:Ham}
\end{eqnarray}

Here, the fermionic operators $c_{j \alpha}$ correspond to conduction 
electrons, hopping between the atomic sites of magnetic ions with spins 
$\vec{S}_i$, and the vector $\vec{\sigma}^{\alpha \beta}$ is composed of Pauli
matrices. Electron concentration is denoted by $x$, hence the hole density
is given by $1-x$ (we note that in the experimental literature on the CMR 
manganates,
the opposite convention is often used). In order to discuss our results 
within the context of experimentally 
observed magnetic phase diagrams of the CMR manganates, we also include the 
antiferromagnetic superexchange coupling $J$ between the ionic spins. We will 
treat the ionic spins as classical, $S \gg 1$; quantum corrections are not 
expected to modify the effects of
Hubbard term in a qualitative way. For the case of CMR manganates, $S=3/2$,
and the band theory calculations\cite{Freeman,Satpathy} suggest the typical 
values of $t\sim0.3-0.5$ eV, $J_H\sim2.5$ eV, and $U \sim 8$ eV.  The value of 
$J$ can be roughly estimated 
from the experimentally observed N\'{e}el temperatures in the fully doped 
(no conduction $e_g$ electrons, $x=0$) case\cite{Schiffer,Martin,Mitchell}, 
$T_N \sim 100-200K$, yielding 
$J \sim 5-10 {\rm meV}$. We will consider the case of square (2D) or simple 
cubic (3D) lattice, assuming that the lattice spacing is equal to unity.  
Throughout the paper, chemical potential is denoted by $\mu - (J_H/2)$.

While the important and highly non-trivial effects of the orbital degree of 
freedom in the CMR manganates are of great interest to both theorists and 
experimentalists working in the field\cite{Tokurabook}, in writing Eq. 
(\ref{eq:Ham}) we 
assumed that there is only one atomic orbital available to conduction 
electrons at each site. The reasons for this drastic simplification are 
three-fold: (i) in the CMR manganates, orbital structure is strongly dependent
on the crystalline environment and varies for different compounds and doping 
levels\cite{tetra}; (ii) yet another situation takes place for magnetic 
semiconductors exhibiting CMR, like EuS or EuSe, where the three $t_{2g}$ 
conduction bands show no Jahn--Teller splitting; 
therefore at present it seems difficult to conclude that a realistic treatment
of orbital effects is crucial for understanding the basic physics of the
CMR phenomenon; and (iii) we will see that the effects of Hubbard repulsion are
rather complex already in the single-orbital case; we believe that these
should be understood before a more complicated model\cite{Marcelo} is
advanced. 

We begin with a brief overview of the low-temperature properties of the double 
exchange model, Eq. (\ref{eq:Ham}) with $U=0$. The carrier spectrum in the
ferromagnetic state of a 2D (3D) system is given by 
$\epsilon^{\uparrow, \downarrow}_{\vec{k}}= \mp (J_H/2)+\epsilon_{\vec{k}}$ 
with $\epsilon_{\vec{k}}=-t\cos k_x-t \cos k_y (-t \cos k_z)$. A sufficiently
large value of Hund's rule coupling $J_H$ then results in a complete spin 
polarisation ({\it half-metal}) of conduction electrons within the entire 
doping range of $0 < x \leq 1$, in agreement with experimental 
data\cite{Park} (see, however, Ref. \cite{Pickett}).  
At $J_H \rightarrow \infty$ and $J=0$, any deviation of ionic spins from 
ferromagnetic order results, via the double exchange mechanism\cite{Anderson},
in a narrowing of the spin-up conduction 
band and therefore costs positive energy. The corresponding value of spin 
stiffness 
$D$ is then proportional to conduction band energy, $DS = |E|/4d$ with 
$E=\int \epsilon_{\vec{k}} n_{\vec{k}} d^dk/(2 \pi)^d$ (where
$n_{\vec{k}}$ is the Fermi distribution function and $d$ is the dimensionality
of the system) and is therefore symmetric in electron density $x$ with 
respect to the quarter-filling, $x=1/2$. In a more general case of finite 
$J_H$ and $J >0$, this {\it double exchange ferromagnetism} competes against 
antiferromagnetic tendencies, which originate from two distinct physical
sources.  In addition to the direct superexchange contribution $J$ (which is
responsible for the antiferromagnetism of the system at $x=0$ and, roughly, 
can be assumed to be doping-independent), there arises an {\it indirect} 
antiferromagnetic interaction\cite{prb03} which further lowers the relative 
energy of antiferromagnetic phases. This interaction is due to virtual 
transitions of
conduction electrons between the two components of the spin-split 
band\cite{formal}, and its strength increases with increasing electron 
density $x$. 
Indeed, the net antiferromagnetic contribution to the spin stiffness $DS$ 
of a double exchange
ferromagnet at $x \ll 1$ equals $-J - \pi x^2 t^2/(2 J_H)$ in 2D and 
$-J-(6 \pi)^{2/3}x^{5/3}t^2 /(10 J_H)$ in 3D, and grows to  $-J-t^2/(4J_H)$ at 
$x=1$. While the actual destabilisation of the 
ferromagnetic phase with increasing
strength of antiferromagnetism proceeds via phase 
separation\cite{Nagaevbook,Dagottobook,jap02}, rather than a spin-wave 
mediated phase transition, this
behaviour of spin stiffness is in line with the overall conclusion on the phase
diagram asymmetry: {\it in the} $U=0$ {\it case, antiferromagnetic tendencies
are more pronounced at} $x>1/2$ {\it than at} $x<1/2$.   

This expected behaviour does not agree\cite{Mishra} with the experimentally 
observed  low-temperature magnetic properties of the CMR 
manganates\cite{Schiffer,Martin,Mitchell}. In broad terms, it is fair to say that {\it the
CMR manganates are ferromagnetic} \cite{accurate} {\it at} $x>1/2$ {\it and 
antiferromagnetic at} $x \leq 1/2$.
Although the presence of narrow ferromagnetic regions (or possibly 
ferro-antiferromagnetic phase separation) at $x<1/2$ has been reported in 
some cases, the wide ferromagnetic area is always located at low hole 
doping, $x>1/2$. The investigation of lightly-doped manganates 
with $1-x \stackrel{<}{\sim} 0.1$ is  complicated by the sample
preparation issues. So far, only the 3D perovskite 
materials are available in this region; these are 
typically\cite{Schiffer,Martin} found to remain ferromagnetic down to the 
very low values of $1-x$, with a likely exception of the $x=1$ 
endpoint\cite{nohole}. This is in contrast with robust 
N\'{e}el antiferromagnetic ordering, characteristic for all manganates at 
$x \ll 1$.

This qualitative discrepancy can be alleviated by taking into account the 
on-site Coulomb interaction, $U$. The latter does not affect the energy nor  
the carrier spectrum of the fully-magnetised half-metallic ferromagnetic 
state; however, 
when the neighbouring spins (say, $\vec{S}_i$ and $\vec S_j$) are out of 
alignment, there arises a 
non-zero hopping matrix element\cite{offdiag}, connecting the conduction 
electron state at site $i$ with the electron spin directed along $\vec{S}_i$, 
and the state at site $j$ with the spin anti-aligned with $\vec{S}_j$. Thus, 
when two electrons are placed on sites $i$ and $j$, there is a non-zero 
quantum-mechanical probability of double occupancy on-site, and the associated
Coulomb energy: the electrons repel each other.  Therefore, {\it Hubbard
interaction enhances ferromagnetism}, and the strength of this effect 
increases with $x$\cite{classical}. Moreover, at a finite temperature $T$, 
when the ionic
spins are misaligned due to thermal fluctuations, {\it an unusual T-dependent
electron-electron repulsion arises in the ferromagnetic phase} (in Sect. 
\ref{sec:scat} below we will see that there also arises another, essentially
many-body, contribution to the effective electron-electron interaction).
 We note that
both of these effects, which we will consider in some detail below, are 
absent in the widespread simplified picture of double exchange, when the
value of Hund's rule coupling is assumed to be infinite, making the double
occupancy impossible.

Although $U$ is, in fact, the largest energy scale in the problem, we will
use the Hartree--Fock approximation, which formally becomes accurate only
at small values of $U/J_H$. It is,
however, expected that substituting larger values of $U$ into our equations
should yield the estimates which are adequate at the qualitative level. We
note that in general, this ``stretching'' of the 
Hartree--Fock scheme requires some
caution, as, for example, the energy of a single spin-down particle
in the ferromagnetic state and in the presence of a partially filled
spin-up band of width $W\ll U$ is clearly of the order of $J_H+W$\cite{single}.
This is in contrast to the  Hartree--Fock result for the energy of the spin 
down electron, 
\begin{equation}
\tilde{\epsilon}^\downarrow_{\vec{k}}=\epsilon^\uparrow_{\vec{k}}+J_S
\,,\,\,\,\,
J_S=J_H+xU
\label{eq:Stoner}
\end{equation}
[here, $J_S$ is the mean-field (Stoner) band splitting].
However, the contribution of the Coulomb energy to the properties studied in
the present paper originates from an integral over {\it many} inter-subband
contributions (or over many spin-down electron states); it is hoped that 
the Hartree--Fock mean field-type 
approximation is more reliable in such a case.   

We will be interested in the experimentally relevant case when the value of 
$J_S$ is large in comparison to  the Fermi energy,  $\epsilon_F$. It should be 
noted that in this regime, the family of models with the 
Hamiltonian (\ref{eq:Ham}) and different values of the ratio $xU/J_H$ provides
a connexion between the conventional double exchange system ($U=0$)
and the large-$U$ Hubbard model ($J_H\rightarrow 0$).  
In fact, due to the considerable uncertainty in the values
of $t$ and $J_H$ quoted in the literature, it is not clear whether $J_H$ alone
could always account for a complete carrier spin-polarisation  
in the ferromagnetic state of the CMR manganates\cite{Pickett}. It is thus 
possible that in
real systems, the half-metallic state (which at the mean-field level is implied
by the condition $J_S> \epsilon_F$) would not have been reached without 
further enhancement of band splitting by the on-site Coulomb 
repulsion $U$ [cf. Eq. (\ref{eq:Stoner})]. If this is indeed the case, it 
might lead to potentially important and novel many-body effects, at both 
zero and finite temperatures. These lie beyond the mean-field approach 
taken in the 
present work -- within the Hartree-Fock scheme based on Eq. (\ref{eq:Stoner}), 
it is indeed unimportant whether  the perceived half-metallicity is partly 
due to the effects of Hubbard repulsion.  


The spin wave theory of double exchange ferromagnets in the  presence of 
on-site Coulomb repulsion is constructed in Section \ref{sec:spinwave}. 
We evaluate the spin stiffness, $D$, and show that the on-site interaction
{\it strengthens ferromagnetism}, while restoring the {\it correct 
asymmetry in the
doping dependence} of $D$. In addition, we show that this interaction 
results in a {\it suppression of magnon energy near the zone boundary}, 
in comparison with the nearest-neighbour Heisenberg dispersion law. 
We then calculate the strength of the novel 
temperature-dependent electron-electron interaction (Sect. \ref{sec:scat}). 
While this interaction appears negligible in the manganates, in case of
lightly doped CMR magnetic semiconductors it does lead to an appreciable 
renormalisation of
nearest-neighbour Coulomb repulsion and (correlated) carrier hopping
amplitude.

The effects of Hubbard repulsion on phase separation in double exchange magnets
at $T=0$ are discussed in Sect. \ref{sec:phase}.  In the $U=0$ case, the 
zero-temperature phase diagram (much like the doping dependence of spin-stiffness) 
suggests that the area of stability of the homogeneous ferromagnetic state
is shifted towards the electron-doped end, $x<0.5$, which is at variance
with generic experimental observations (see above). We show that inclusion of
$U$ alleviates this difficulty as well.

The observed suppression of the carrier density of
states near the Fermi level at low temperatures\cite{Mitra} is likely to be
of the same origin as the much more pronounced depletion of the density 
of states\cite{Dessau,Biswas} (sometimes termed ``pseudogap'') 
in the vicinity of the Curie 
temperature. It appears, in turn, that in order to adequately describe 
the physics of CMR phenomenon one has to understand the nature of the 
pseudogap. It is therefore important that a proper description
of a low-temperature ferromagnetic state of the CMR manganates should
include the correct energy dependence of the density of states.
In Sect. \ref{sec:dos} we show that standard Altshuler--Aronov
mechanism utilizing a combination of  Coulomb repulsion
(or the effective electron-electron interaction
derived in Sect. \ref{sec:scat}) and impurity scattering cannot account
for the measured depletion of the density of states. This signals 
the insufficiency of our mean-field treatment in this case, and the
presence of strong energy-dependent correlation effects even at low
temperatures.

The implications of our findings are further summarised in Sect. 
\ref{sec:conclu}, where we also discuss prospective directions for the 
future theoretical and experimental work in the field. On the whole, 
our results indicate that the effects of the Hubbard $U$ 
are indeed crucial for the  
understanding of magnetic properties of the CMR compounds. Qualitatively,
this suggests that in addition to the familiar double-exchange band-narrowing
effects, the {\it correlated behaviour of spin-polarised large-$U$ Hubbard 
carriers}
should be recognised as an important mechanism underlying the physics
of CMR compounds.

\section{SPIN WAVE THEORY}
\label{sec:spinwave}

Low-temperature spin-wave properties of double exchange ferromagnets with
$U=0$ are well understood (see Ref. \cite{prl00} and references therein).
In parallel with Ref.  \cite{prl00}, we begin our treatment of the Hamiltonian,
Eq. (\ref{eq:Ham}) with $U \neq 0$, with the standard Holstein--Primakoff 
transformation, followed by a canonical transformation of the form
\begin{eqnarray}
{\cal H} & \rightarrow& {\cal H}^\prime=\exp(-W) {\cal H} \exp(W)\,, \nonumber \\
W&=&\frac{J_H}{\sqrt{2SN}} \sum_{\vec{k},\vec{p}} \left(W_{\vec{k},\vec{p}}
c^\dagger_{\vec{k} \uparrow} c_{\vec{k}+\vec{p}\downarrow} a^\dagger_{\vec{p}} -
{\rm H.c.} \right)\,,
\label{eq:canon}
\end{eqnarray}
where $a^\dagger_{\vec{p}}$ is the magnon creation operator, and 
$N$ is the total number of  lattice sites. The resulting Hamiltonian, 
${\cal H}^\prime$, takes form of a series in powers of $1/\sqrt{S} \ll 1$,
with the leading-order term,
\begin{eqnarray}
{\cal H}^\prime_0&=&\sum_{\vec{k},\alpha} \epsilon^\alpha_{\vec{k}} 
c^\dagger_{\vec{k}\alpha} c_{\vec{k}\alpha}+ U \sum_{1 \div 4}^{}{\!\!\,}^{'} 
c^\dagger_{1\uparrow} c^\dagger_{2\downarrow}c_{3\downarrow}c_{4\uparrow}+dNJ
\label{eq:H0}
\end{eqnarray}
(here, $\sum^\prime$ means that the quasimomentum conservation law is obeyed). Since
we will be interested in the leading-order (classical) spin wave properties, 
we will need only the two further terms in this series. Here, in addition to 
the ``usual''
terms occurring already in the non-interacting model\cite{prl00},
\begin{eqnarray}
{\cal H}^\prime_1&=&\frac{J_H}{\sqrt{2SN}}\sum_{\vec{k},\vec{p}}\left\{\left[
W_{\vec{k},\vec{p}}(\epsilon^\uparrow_{\vec{k}}-\epsilon^\downarrow_{\vec{k}+
\vec{p}}) -1 \right] \times \right. \nonumber \\
&&\left. \times c^\dagger_{\vec{k}\uparrow} c_{\vec{k}+\vec{p} \downarrow} a^\dagger_p + {\rm H. c.} \right\}, 
\label{eq:H1}\\
{\cal H}^\prime_2&\stackrel{\rm eff}{=}&-\frac{J^2_H}{4SN}
\sum_{1\div 4}^{}{\!\!\,}^{'} \left\{W_{1,3} W_{2,4}\left( \epsilon^\uparrow_1 -
\epsilon^\downarrow_{1+3}+\epsilon^\uparrow_2-
\epsilon^\downarrow_{2+4}\right)- \right. \nonumber \\
&&- 
\left.2W_{1,3} -2W_{2,4}-\frac{2}{J_H}\right\} c^\dagger_{1\uparrow}
c_{2\uparrow} a^\dagger_3 a_4 - \nonumber \\
&&-\frac{2J}{S}\sum_{\vec{k}}\left(d+
\frac{1}{t}\epsilon_{\vec{k}}\right)a^\dagger_{\vec{k}}a_{\vec{k}}\,
\label{eq:H2}
\end{eqnarray}
(where $W_{1,3}$ stands for $W_{\vec{k}_1,\vec{k}_3}$ etc., and the sum in the
first term is over $\vec{k}_1$,...,$\vec{k}_4$),
we find two interaction-induced terms,
\begin{eqnarray}
{\cal H}^\prime_{i1}&\stackrel{\rm eff}{=}&\frac{UJ_H}{\sqrt{2S}N^{3/2}} 
\sum_{1\div 5}^{}{\!\!\,}^{'} \left\{W_{1,5}c^\dagger_{1\uparrow}c^\dagger_{2\uparrow}
c_{3\downarrow}c_{4\uparrow}a^\dagger_5+ {\rm H. c.}\right\},\,
\label{eq:Hi1}\\
{\cal H}^\prime_{i2}&\stackrel{\rm eff}{=}& \frac{UJ_H^2}{4SN^2}
\sum_{1\div 6}^{}{\!\!\,}^{'}\left\{W_{1,5}W_{4,6}-W_{2,5}W_{4,6}\right\} 
\times \nonumber \\ && \times c^\dagger_{1\uparrow}c^\dagger_{2\uparrow}
c_{3\uparrow}c_{4\uparrow}a^\dagger_5a_6\,. 
\label{eq:Hi2}
\end{eqnarray}
In writing Eqs. (\ref{eq:H2}--\ref{eq:Hi2}), we omitted the terms containing
more than one spin-down fermion operator $c_{\vec{k}\downarrow}$ or 
$c^\dagger_{\vec{k}\downarrow}$ (hence the ``eff'' above the equality signs).
Due to the absence of spin-down electrons
in the half-metallic ferromagnetic ground state, these terms will not 
contribute to the quantities which are of interest to us here.

We find it advantageous to choose the coefficients $W_{\vec{k},\vec{p}}$
in such a way that the leading-order single-particle electron-magnon 
scattering, 
${\cal H}^\prime_1$, is cancelled  by the average contribution of
${\cal H}^\prime_{i1}$ [see Fig. \ref{fig:diag} {\it (a)}]:
\begin{eqnarray}
1-\left(\epsilon^\uparrow_{\vec{k}}-\epsilon^\downarrow_{\vec{k}+\vec{p}}
\right)W_{\vec{k},\vec{p}}&=&-UxW_{\vec{k},\vec{p}}+\frac{U}{N}\sum_{\vec{q}}
n_{\vec{q}}W_{\vec{q},\vec{p}}\,,
\label{eq:condW}\\
n_{\vec{k}}\equiv\langle c^\dagger_{\vec{k} \uparrow}c_{\vec{k} 
\uparrow} \rangle\,,\,\,\,\,\,x&=&\frac{1}{N}\sum_{\vec{k}}n_{\vec{k}}\,.
\end{eqnarray}

Within the mean-field picture, condition 
(\ref{eq:condW}) implies that the average number of magnons with a given 
momentum $\vec{p}$,
\begin{equation} 
{\cal N}_{\vec{p}}=\langle a^\dagger_{\vec{p}} a_{\vec{p}} \rangle\,,
\label{eq:magnonnumber}
\end{equation}
remains constant\cite{highercoll}. In other words, it represents the
optimal choice in separating the two distinct branches of excitations 
(magnons and electrons/holes). 

Equation (\ref{eq:condW}) is solved by

\begin{equation}
W_{\vec{k},\vec{p}}=\frac{R_{\vec{p}}}{\epsilon^\uparrow_{\vec{k}}-
\epsilon^\downarrow_{\vec{k}+\vec{p}}-Ux}\,,\,\,\frac{1}{R_{\vec{p}}}=
1+\frac{U}{N}\sum_{\vec{q}}\frac{n_{\vec{q}}}{\epsilon^\uparrow_{\vec{q}}-
\epsilon^\downarrow_{\vec{q}+\vec{p}}-Ux}\,,
\label{eq:W}
\end{equation}
which at $U\rightarrow 0$ reduces to the familiar form\cite{Nagaev98}, 
used earlier for the non-interacting case\cite{prl00,Nagaev98}. 

These expressions can be further simplified in the experimentally 
relevant case of $J_S \gg \epsilon_F$, where $\epsilon_F=\mu+td$ is the Fermi energy,
measured from the bottom of the spin-up subband. In particular, we then find
\begin{equation}
\frac{J_S}{R_{\vec{p}}} \approx J_H -\frac{U}{NJ_S}\sum_{\vec{q}}
n_{\vec{q}} (\epsilon_{\vec{q}}-\epsilon_{\vec{q}+\vec{p}})=
J_H+\frac{|E|U}{J_S} (1+\frac{\epsilon_{\vec{p}}}{td})\,.
\label{eq:RplargeJS}
\end{equation}
If we also restrict ourselves to the case of small magnon momenta, 
$p,p' \ll 1$, the second term on the r.\ h.\ s. of Eq. (\ref{eq:RplargeJS})
can be omitted, and we obtain

\begin{eqnarray}
{\cal H}^\prime_1 &\approx&\frac{Ux}{J_S \sqrt{2SN}} \sum_{\vec{k}, \vec{p}}
\left\{\vec{v}_{\vec{k}} \cdot \vec{p} c^\dagger_{\vec{k} \uparrow} 
c_{\vec{k}+\vec{p} \downarrow} a^\dagger_{\vec{p}} + {\rm H. c.} \right\}\,, 
\label{eq:H1smallp}\\
{\cal H}^\prime_{i1} &\approx&\frac{U}{2 \sqrt{2S}N^{3/2}J_S} 
\sum_{1\div 4, \vec{p}}^{}{\!\!}^{'}
 \left\{ (\vec{v}_1-\vec{v}_2)\cdot \vec{p} \times \right. \nonumber \\
&&\left.\times
c^\dagger_{1 \uparrow} c^\dagger_{2 \uparrow}c_{3 \downarrow} 
c_{4 \uparrow} a^\dagger_{\vec{p}} + {\rm H. c.} \right\}\,, 
\label{eq:Hi1smallp}\\
{\cal H}^\prime_{i2} &\approx&\frac{U}{8SN^2J_S^2} 
\sum_{1\div 4, \vec{p}, \vec{p}\,'}^{}{}^{'} \left\{ 
(\vec{v}_2-\vec{v}_1)\cdot \vec{p} \right\} \left\{
(\vec{v}_3-\vec{v}_4)\cdot \vec{p}\,\,' \right\} \times \nonumber \\
&&\times
c^\dagger_{1 \uparrow} c^\dagger_{2 \uparrow}c_{3 \uparrow} 
c_{4 \uparrow} a^\dagger_{\vec{p}} a_{\vec{p}\,'}\,,
\label{eq:Hi2smallp}
\end{eqnarray}
where $\vec{v}_{\vec{k}} \equiv \partial \epsilon_{\vec{k}}\,/
\partial\vec{k}$ is the electron velocity. 

The spin-wave energy, $\omega_{\vec{p}}$, is equal to magnon 
self-energy\cite{leading}, which
in turn can be evaluated perturbatively (in $1/\sqrt{S}$). In addition to the
first-order contributions from ${\cal H}^\prime_2$ and ${\cal H}^\prime_{i2}$,
there is a number of second-order corrections from ${\cal H}^\prime_1$ and 
${\cal H}^\prime_{i1}$. Owing to the condition (\ref{eq:condW}), these
second-order terms cancel each other, with the sole exception shown 
diagrammatically in Fig. \ref{fig:diag} {\it (b)}. In drawing and evaluating 
this diagram,
we make a drastic simplification of a mean-field type, corresponding to the
Hartree decoupling of the interaction term in Eq. (\ref{eq:H0}). Namely,
we do not include any spin-up -- spin-down electron vertices and use the
expression,
\begin{equation}
G_\downarrow(\omega, \vec{k}) = 1/(\omega -{\epsilon}_{\vec{k}}-
J_S+\mu + i0 \cdot {\rm sign \omega})\, 
\label{eq:Greendown}
\end{equation}
for the spin-down electron Green's function. The Green's function for a spin-up electron in the
half-metallic case is given by the usual formula,
$  G_\uparrow(\omega, \vec{k}) = 1/(\omega -{\epsilon}_{\vec{k}}+\mu + 
i0 \cdot {\rm sign \omega})$\,. 

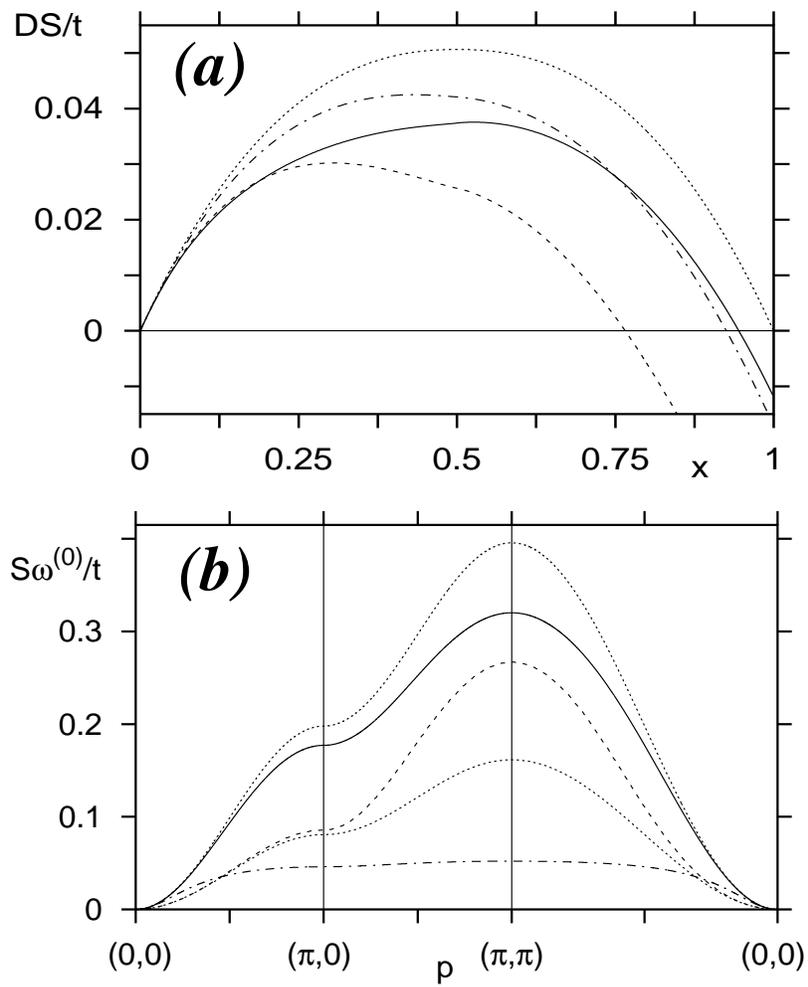
\begin{figure}
\caption{ {\it (a)} Doping dependence of spin stiffness for a two-dimensional 
system with $J=0$, $J_H/t=5$ (solid and dashed lines, corresponding to
$U/t=16$ and $U=0$ cases, respectively). The dashed-dotted line corresponds to 
$J_H/t=14.6$
and $U=0$, and the dotted line -- to $J_H \rightarrow \infty$.
{\it (b)} The leading-order  magnon 
energy in a 2D system with $U/t=16$, $J=0$, and $x=0.6$. Solid, dashed 
and the upper dotted lines are plotted using Eq. (\ref{eq:magnonspec}) and 
correspond to $J_H/t=5$, $J_H/t=0.2$, and 
$J_H \rightarrow \infty$, respectively. The dashed line is the 
result\protect\cite{Nagaev69} 
for $J_H/t=5$, $U=0$, and the lower dotted line is
the corresponding Heisenberg fit.  }
\label{fig:stiff}
\end{figure}

The resultant expression for long-wavelength magnon dispersion takes the
usual form, $\omega_{\vec{p}} = Dp^2$, where the spin-stiffness $D$ is given by
\begin{equation}
DS=\frac{|E|}{4d}-J-
\frac{U^2 x(1-x)+J_S^2}{2dJ_S^3} \int 
n_{\vec{k}}v^2_{\vec{k}} \frac{d^dk}{(2 \pi)^d}\,.
\label{eq:stiff}
\end{equation}
The doping dependence of the spin-stiffness D for a two-dimensional system is
illustrated in Fig. \ref{fig:stiff} {\it (a)}. Here, the solid line 
shows our result, Eq. (\ref{eq:stiff}), for $J=0$ and experimentally
relevant values $J_H/t=5$, $U/t=16$. The effect of Hubbard repulsion 
becomes clear from comparison with the dashed line, corresponding to the
$J_H/t=5$, $U=0$ case \cite{exact}. We see that in the presence of the
on-site repulsion, the magnitude of $D(x)$ increases, and
the maximum is shifted towards $x>0.5$. The dashed-dotted line 
corresponds
to the $J_H/t=14.6$, $U=0$ case, which is characterised by the same value
of mean-field band-splitting $J_S$ as $J_H/t=5$, $U/t=16$ system at the 
experimentally important value of electron density, $x=0.6$. While  
numerically the two results at $0.5<x<0.8$ are relatively close, the 
dashed-dotted line possesses a larger slope, and still reaches maximum below
quarter-filling, $x=0.5$. For completeness, we note that the classical 
$J_H \rightarrow \infty $ result
(dotted line) is symmetric and corresponds to the largest magnitude of $D(x)$. 
We conclude that at a finite $J_H$, in addition to an overall increase in $D$, 
{\it the inclusion of Hubbard repulsion leads to a
relative increase of spin stiffness at $x>1/2$}, which is consistent with
the experimental observation that the ferromagnetic tendencies in the CMR
manganates are more pronounced in this doping region.
We note that an earlier mean field study\cite{Kieinert} of the effects of the
Hubbard repulsion on the Curie temperature, $T_C(x)$, suggested somewhat
similar trends.

The overall increase of spin stiffness originates from the mean-field effects
discussed in Sect. \ref{sec:intro} and corresponds to substitution
$J_H \rightarrow J_S$ in the last term of  Eq.(\ref{eq:stiff}). Since $J_S>J_H$, the
(negative) pre-factor in front of the (positive) integral decreases in
comparison with the $U=0$ case, resulting in the increase of $D$. 
At sufficiently low values of $x$, however, this tendency is counter-balanced 
by {\it effects of the Hubbard correlations}, which contribute the 
quantity $U^2x(1-x)$ to the numerator of this pre-factor. The underlying
physics will be discussed in Sect. \ref{sec:scat} below; here we merely note
that as a result, the spin stiffness well below the quarter-filling, 
$x<0.2$, may actually be 
somewhat suppressed in comparison with the $U=0$ value. For 
the case of $J_H=5t$, $U=16t$, this takes place for $x<0.19$ [see Fig.
\ref{fig:stiff} {\it (a)}]. 

As for a quantitative comparison of our results for $D(x)$ at $x>0.5$ with
the experimental data for the CMR manganates, 
this appears problematic due to a number of reasons:
(i) available experimental results on the doping dependence of 
spin-stiffness\cite{stiffexp}
are still incomplete; (ii) it is known\cite{prl00} that quantum corrections,
not included in Eq. (\ref{eq:stiff}), lead to an appreciable renormalisation of
spin stiffness magnitude; and (iii) the values of band (and orbital) structure 
parameters and
direct exchange integrals for particular compounds are known with a large
degree of 
uncertainty. Nevertheless, it is adequate to say that qualitatively, both the 
overall profile and the magnitude of spin-stiffness, as given by 
Eq. (\ref{eq:stiff}), are consistent with the experimental results for 
$D(x)$  within the metallic ferromagnetic region, $0.2<1-x<0.5$.

When  spin stiffness [as shown in Fig. \ref{fig:stiff} {\it (a)} for $J=0$]  
turns negative, the ferromagnetic ground
state can only be stabilised by including a sufficiently strong direct
ferromagnetic exchange coupling, $J<0$. On the contrary, a positive value of
spin stiffness does {\it not} guarantee the stability of a uniform 
ferromagnetic
ground state, since the latter might still be unstable with respect 
to phase separation (see Sect. \ref{sec:phase} below).

Within the mean-field approach taken here, expression (\ref{eq:stiff}) is 
expected to hold at $J_S \gg \epsilon_F$ for all values of the ratio $U/J_H$, 
except for the $J_H \rightarrow 0$ case of a pure Hubbard model. In this case,
the ionic spins are fully decoupled from the itinerant ones, and the 
leading-order (in $1/S$) term in the magnon energy vanishes. Formally, 
at $J_H \rightarrow 0$ the second term on the r.\ h.\ s. of 
Eq. (\ref{eq:RplargeJS}) cannot be omitted 
even for small $p$, and Eqs. (\ref{eq:H1smallp}--\ref{eq:stiff}) 
become invalid.  
We note that if the value of $U$ is sufficiently large,
the conduction electrons may still be in the fully spin-polarised 
ferromagnetic state as expected for a partially-filled Hubbard model
below half-filling, $x<1$. While this is always the case within
the present mean field treatment, actual identification of the stability
region for a ferromagnetic state of a large-$U$ Hubbard model remains an 
open problem\cite{Wurth,Wurth2}. 
Although this subject is well beyond the scope of 
the present work, it is important to note that (i) it is likely that over 
a broad range of doping values, the instability of the fully spin-polarised 
state of the Hubbard model  ($J_H=0$) results
only in a {\it partial} reduction of magnetisation\cite{Wurth}, and 
(ii) it is possible that
allowing for a small but finite value of $J_H$ greatly enhances stability
of the fully spin-polarised state (cf. Ref. \cite{Wurth2}).

In view of relatively small values of $J_H$, reported in the bandstructure 
calculations, it is important to study the crossover to the free-spin
($J_H \rightarrow0$) case in some detail. The leading-order (in both 
$\epsilon_F/J_S$ 
and $1/S$) term in the magnon energy originates from ${\cal H}_2^\prime$ and 
in the absence of a direct coupling $J$ has the form 

\begin{equation}
\omega^{(0)}_{\vec{p}}=\frac{|E|}{2S}\left(1+\frac{\epsilon_{\vec{p}}}{td}
\right)
\frac{J_H^2+xJ_H \frac{|E|U^2}{J_S^2}\left(1+
\frac{\epsilon_{\vec{p}}}{td}\right)}{\left[J_H+\frac{|E|U}{J_S}\left(1+
\frac{\epsilon_{\vec{p}}}{td}\right)\right]^2}\,.
\label{eq:magnonspec}
\end{equation}
Here, momentum $\vec{p}$ is allowed to take any value within the Brillouin 
zone. When $J_H$ is sufficiently small, the second term in the denominator
dominates\cite{opposite}, 
provided that $p^2\stackrel{>}{\sim}(J_SJ_H)/(U|E|)$. The magnon
energy then saturates at a constant value, $\omega_{\vec{p}} = xJ_H/(2S)$,
which is consistent with a physical picture of independent ionic spins 
$\vec{S}_i$ subject to an effective magnetic field of the magnitude $xJ_H/2$.
The latter is created by the rigid ferromagnetic Fermi sea of the large-$U$
Hubbard carriers\cite{reduced}. 
This situation is illustrated by a dashed-dotted line in Fig.
\ref{fig:stiff} {\it (b)}, corresponding to $J_H/t=0.2$, $U/t=16$ and $x=0.6$.
While for the experimentally relevant value of $J_H/t=5$ (solid line) 
spin-wave energy does not reach saturation, the effects of 
suppression of the magnon energy (in comparison with the pure cosine 
Heisenberg law -- see the upper dotted line, corresponding to 
$J_H \rightarrow \infty$) are still felt near the zone boundary\cite{corr}. 
These become more pronounced (possibly leading even to a local minimum of 
spin-wave dispersion at the point $\vec{p}=\{\pi,\pi\}$), if a direct 
antiferromagnetic 
coupling $J>0$ between the ionic spins is taken into account. The latter
gives rise to an extra term, $-2J(\epsilon_{\vec{p}}+td)/(tS)$, which should be 
added to the r.\ h.\ s. of Eq. (\ref{eq:magnonspec}). Thus,
{\it at the moderate values of $J_H$}  within the experimentally relevant
range, the {\it spin-wave dispersion in a double exchange ferromagnet with a 
large
on-site repulsion $U$ shows zone-boundary ``softening''} in comparison with 
the nearest-neighbour Heisenberg dispersion law. There has been an extensive
theoretical effort\cite{theosoft,Sun} directed at understanding this property, 
which is
observed experimentally\cite{expsoft} in many (but not all\cite{expnosoft})
CMR compounds. It is important that this 
generic feature is recovered within the present model, as suggested by earlier
variational studies\cite{Sun} of the effects of $U$ on the magnon dispersion.

We note that this zone-boundary softening effect occurs in both 2D and 3D, and 
is entirely due to the Hubbard repulsion, $U$. Indeed, for any dimensionality 
$d$ it can be shown\cite{jap00} that at $U=0$ and $x>0.5$, the magnon 
dispersion at sufficiently large $J_H\stackrel{>}{\sim} \epsilon_F$ 
{\it hardens} towards the zone boundary. This is 
illustrated by the dashed line in Fig. \ref{fig:stiff} {\it (b)}, representing
the $J_H=5t$, $U=0$ case\cite{Nagaev69}. One can see that at large momenta, the
corresponding Heisenberg dispersion law, 
$\omega_{\vec{p}}=2D(td+\epsilon_{\vec{p}})/t$ with the appropriate value of 
spin-stiffness $D$, indeed yields lower magnon energies (lower dotted line).

\section{EFFECTIVE ELECTRON-ELECTRON INTERACTION}
\label{sec:scat}

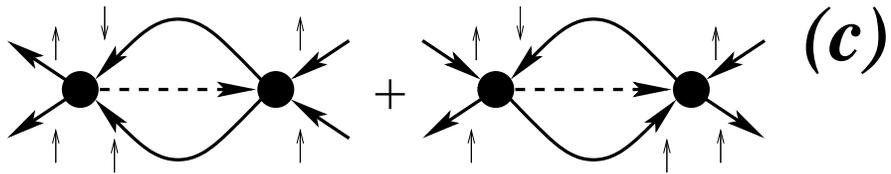
\begin{figure}
\caption{{\it (a)} Schematic representation of Eq.(\ref{eq:condW}), with the 
two vertices corresponding to ${\cal H}^\prime_1$ and  ${\cal H}^\prime_{i1}$
[see Eqs. (\ref{eq:H1}) and (\ref{eq:Hi1})], respectively.
{\it (b)} Second-order interaction correction  to the magnon energy [cf. Eq. 
(\ref{eq:stiff})]. {\it (c)} Second-order contribution to the
temperature-dependent interaction $\Gamma^{\uparrow \uparrow}$ between two
spin-up electrons, see Eq. (\ref{eq:Gamma}). In all cases, solid  and dashed 
lines 
correspond to electron and magnon Green's functions, respectively. Up- and 
down-arrows denote spin of the electrons. When evaluating these diagrams, one
should ensure the proper antisymmetrisation of the the spin-up 
electronic ``legs'' 
of each vertex, taking into account the appropriate momentum dependence. 
Momentum integration is greatly simplified in the large-$J_S$ case considered 
here.}
\label{fig:diag}
\end{figure}

The original on-site Coulomb repulsion, as represented by the last term in
Eq. (\ref{eq:Ham}), acts between electrons with anti-aligned spins. As the
average number of spin-down electrons in a half-metallic double exchange
ferromagnet at low temperatures is negligible, it might seem that
the spin-up electrons remain non-interacting even in the presence of the
Hubbard $U$. However,
as already discussed in Sect. \ref{sec:intro}, at finite temperatures the 
on-site repulsion 
also gives rise to an interaction between electrons with the same sign of
spin projection. The presence of this novel interaction, 
$V_{eff}$, is clear, 
{\it e.g.},  from the form of the operator ${\cal H}^\prime_{i2}$ [see Eq.
(\ref{eq:Hi2})], which at finite $T$ can be averaged over the
equilibrium magnon distribution. Another contribution\cite{order} to 
$V_{eff}$ originates from the second-order processes, involving
various combinations of terms from 
${\cal H}^\prime_{1}$ and ${\cal H}^\prime_{i1}$, 
Eqs. (\ref{eq:H1}) and (\ref{eq:Hi1}). As in Sect. \ref{sec:spinwave} 
above, the condition (\ref{eq:condW}) leads to a massive cancellation among
these second-order diagrams [cf. Fig. \ref{fig:diag} {\it (a)}], 
with the only two surviving terms shown in Fig. \ref{fig:diag} {\it (c)}.

In Fig. \ref{fig:diag} {\it (c)}, vertices correspond to the magnon-electron
interaction, ${\cal H}^\prime_{i1}$, and the solid lines are 
finite-temperature electron Green's functions, 
\begin{equation}
{\cal G}_\uparrow^{-1}(\zeta, \vec{p})=i \zeta-\epsilon_{\vec{p}}+\mu\,,\,\,\,
{\cal G}_\downarrow^{-1}(\zeta, \vec{p})=i \zeta-\epsilon_{\vec{p}}-J_S+\mu \,,
\label{eq:Greens}
\end{equation}
where we again used the Hartree mean-field form for ${\cal G}_\downarrow$ [cf.
Eq. (\ref{eq:Greendown})]. At
low temperatures, $T \ll D$, only the long-wavelength magnons are present,
and the magnon Green's function can be written as 
${\cal G}_m^{-1} \approx i \zeta -D p^2$. Here, the spin stiffness $D$ can be 
evaluated with the help of  
Eq. (\ref{eq:stiff}) or taken directly from the low-temperature neutron 
scattering measurements. Furthermore, at low temperatures Eqs. 
(\ref{eq:Hi1smallp}--\ref{eq:Hi2smallp})
may be used in place of Eqs. (\ref{eq:Hi1}--\ref{eq:Hi2}).

To leading order in $\epsilon_F/J_S$ and $1/S$, the net result for the vertex function,
$\Gamma^{\uparrow\uparrow}_{12,43}$, of 
two spin-up electrons scattering is given by the expression
\begin{eqnarray}
&&\frac{2}{SJ^2_S} \int \frac{d^dp}{(2 \pi)^d} {\cal N}_{\vec{p}}
\left[U(\vec{v}_1 \cdot \vec{p}\,) 
(\vec{v}_4\cdot \vec{p}\,)- \frac{U^2(x-1)}{i (\zeta_1+\zeta_2) -J_S}
\times \right. \nonumber \\
&&\times (\vec{v}_1 \cdot \vec{p}\,) (\vec{v}_4\cdot \vec{p}\,)
 -\frac{U^2}{i(\zeta_1-\zeta_3)-J_S} \times \nonumber \\
&& \left. \times \int \frac{n_{\vec{k}}d^dk} 
{(2 \pi)^d} \left\{(\vec{v}_1 -\vec{v}_{\vec{k}}\,)\cdot \vec{p}\,\right\} 
\left\{(\vec{v}_4 -\vec{v}_{\vec{k}}\,)\cdot \vec{p}\,\right\}\right]\,,
\label{eq:Gamma}
\end{eqnarray}
which should be anti-symmetrised\cite{antisim} with respect to the 
velocities and 
frequencies of the outgoing (incoming) electrons, $\vec{v}_{1,2}\equiv
\partial \epsilon_{\vec{p}_{1,2}} /\partial \vec{p}_{1,2}$ and $\zeta_{1,2}$
($\vec{v}_{3,4}$ and $\zeta_{3,4}$). 
In Eq. (\ref{eq:Gamma}), 
the first term in brackets represents the first-order
contribution of the operator ${\cal H}^\prime_{i2}$, Eq. (\ref{eq:Hi2smallp}),
whereas the other two terms come from the two diagrams shown in Fig.
\ref{fig:diag} {\it (c)};
${\cal N}_{\vec{p}}=[\exp (D p^2/T) -1 ]^{-1}$ 
is the average magnon occupation number, Eq. (\ref{eq:magnonnumber}).

The retardation (frequency-dependence) effects in 
$\Gamma^{\uparrow \uparrow}$ become noticeable only
on a very large electron energy (frequency) scale\cite{retard} 
of $|\zeta_i| \sim J_S$, and
can be omitted whenever only electrons with energies near the Fermi level
are considered. In this case, the effect of electron-electron scattering as
described by the vertex $\Gamma^{\uparrow \uparrow}$ is equivalent to that
of {\it an effective electron-electron interaction} of the form 
\begin{eqnarray}
V_{eff}&=& - \frac{U[J_S+U(2x-1)]}{8dSJ_S^3N} \int p^2 {\cal N}_{\vec{p}}
\frac{d^dp}{(2 \pi)^d}\times\nonumber \\
&&\times
\sum_{1\div4}^{}{\!}^{'}
(\vec{v}_1-\vec{v}_2)\cdot (\vec{v}_3-\vec{v}_4)
c^\dagger_{1\uparrow}c^\dagger_{2\uparrow}c_{3\uparrow} c_{4\uparrow} \,,
\label{eq:Veff}
\end{eqnarray}
where we also used the fact that the long-wavelength magnon dispersion is 
isotropic.
In the case of small $U \ll J_H, \epsilon_F$, our mean-field results, 
Eqs. (\ref{eq:Gamma}) and (\ref{eq:Veff}), can be re-derived within the 
perturbation theory in $U$. As expected on physical grounds, 
$V_{eff}$ vanishes also in the 
$U\rightarrow \infty$ limit, when the double occupancy on-site is forbidden.
The momentum integral occurring on the r.\ h.\ s. of Eq. 
(\ref{eq:Veff}) can be easily evaluated,
\begin{equation}
I(T)\equiv\int p^2 {\cal N}_{\vec{p}}
\frac{d^dp}{(2 \pi)^d} =\left\{ \begin{array}{ll}
\frac{3\zeta(5/2)}{16 \pi^{3/2}} \left(\frac{T}{D} \right)^{5/2} & 
\mbox{in 3D,}\\ & \\\frac{\pi}{24} \left(\frac{T}{D}\right)^2 & \mbox {in 2D,}
\end{array} \right.
\label{eq:magnonintegral}
\end{equation}
where $\zeta(5/2)\approx 1.34$ is the Riemann's {\it zeta}-function. 
We note that the quantity (\ref{eq:magnonintegral}) can also be 
expressed macroscopically as the thermal average of 
\begin{equation}
\frac{1}{2S}\sum_{\alpha=1}^3 \left\{\vec{\nabla} M^\alpha 
\cdot \vec{\nabla} M^\alpha \right\}\,,
\label{eq:macro}
\end{equation}
where $M^\alpha$ are the three components of local 
magnetisation\cite{macro}, $\vec{M}$. This shows
that the appearance of $V_{eff}$ is indeed a direct consequence of the 
misalignment
of neighbouring spins (which in turn is due to the thermal fluctuations; 
cf. Sect. \ref{sec:intro}). 

Although the precise form of $V_{eff}$ in Eq. (\ref{eq:Veff}) 
obviously has only a mean-field validity,
we emphasise that qualitatively this effect, which has a clear physical 
origin, will survive in an exact treatment. If anything, the mean-field 
approach
yields a smaller magnitude of $V_{eff}$: indeed, Eqs. (\ref{eq:Greens})
overestimate the energy of a spin-down electron, which enters into the 
denominators of diagrammatic expressions in Fig. \ref{fig:diag} {\it (c)}.

Thus, we conclude that {\it with increasing temperature T, there arises an
effective interaction between the spin-polarised carriers in a double
exchange ferromagnet} with $U >0$.  
For the purposes of order-of-magnitude estimates, one can assume that
electron dispersion is isotropic, $\vec{v} \approx \vec{p}/m_*$, where
$m_*$ is an effective mass of electron (or hole). The interaction, 
Eq. (\ref{eq:Veff}),
then takes form of a simple $p$-wave scattering,
\begin{equation}
V_{eff} = \frac{2}{Nm_*^2} \sum_{\vec{s}, \vec{q}, 
\vec{q}^{\,\prime}} V(T)(\vec{q} \cdot\vec{q}^{\,\prime}) 
c^\dagger_{\vec{s}+\vec{q}^{\,\prime} \uparrow}
c^\dagger_{\vec{s}-\vec{q}^{\,\prime} \uparrow}
c_{\vec{s}-\vec{q} \uparrow}c_{\vec{s}+\vec{q} \uparrow}\,,
\label{eq:Veffsmallk}
\end{equation}
with 
\begin{equation}
V(T) = \frac{U[J_S+U(2x-1)]}{4dSJ_S^3}
I(T)\,.
\label{eq:pwaveampl}
\end{equation}

In real space, the effective interaction, Eq. (\ref{eq:Veff}), takes form
\begin{equation}
V_{eff}=\frac{t^2}{2}V(T) \sum_{i,\Delta} \left\{ c^\dagger_{i \uparrow}
c^\dagger_{i+\Delta \uparrow}c_{i+\Delta \uparrow} c_{i \uparrow}
-c^\dagger_{i+\Delta \uparrow}
c^\dagger_{i \uparrow}c_{i \uparrow} c_{i-\Delta \uparrow} \right\}\,.
\label{eq:Vreal}
\end{equation}
Here, for each lattice site $i$ a summation over its $2d$ nearest neighbours
(labeled $i+\Delta$) is performed. The first term in Eq.(\ref{eq:Vreal}) 
contains a product of carrier densities; it renormalises the ``bare'' 
repulsion between electrons on the neighbouring sites, which is present 
in reality but not included in our model, Eq. (\ref{eq:Ham}). In the case
of the CMR manganates, this bare repulsion
is expected \cite{Satpathy2002} to be of the order of $V_{nn} 
\stackrel{<}{\sim} 0.05$ eV; 
on the other hand,
the value of $t^2V(T)$ as found from Eq. (\ref{eq:pwaveampl}) with $U= 8$ eV, 
$t=0.5$ eV, $J_H=2.5$ eV and $x=0.6$ is $t^2V(T) \sim   10^{-4} 
(T/T_C)^{5/2}$ (in units of eV, for the 3D case; we also assumed that the Curie
temperature $T_C$ is of the order of the zero-temperature spin-stiffness, 
$D$). We thus see that in the manganates, the contribution of $V_{eff}$, Eq. 
(\ref{eq:Vreal}),
to the nearest-neighbour repulsion remains negligible even for $T \sim T_C$.

The situation is likely to be very different for non-manganate magnetic
semiconductors exhibiting CMR. It is expected\cite{Umehara,Steeneken} 
that in the lightly doped ($x \sim 10^{-4}$) EuSe , EuTe, EuO or undoped 
(semimetallic) ${\rm EuB}_6$ the parameters
of the Hamiltonian, Eq. (\ref{eq:Ham}) can be roughly estimated as $U=7$ eV, 
$J_H = 0.4 eV$ and $t=0.5$ eV. Taking also into account the magnitude
of Eu spin (7/2), we find $t^2V(T) \sim -0.2 (T/T_C)^{5/2}$ in units of eV.
Thus, the effects of temperature-dependent renormalisation of $V_{nn}$ in
these compounds may be appreciable.

We note that in the latter example, the sign of $V(T)$ is negative, 
corresponding to an effective {\it attraction} between electrons on the 
neighbouring sites. This clearly contradicts the simple  physical 
picture outlined in Sect. \ref{sec:intro} (based on the small-$U$ 
perturbative considerations).  Indeed, as can be seen from 
Eq. (\ref{eq:pwaveampl}), at sufficiently large values of
$U$ the effective interaction can in fact become attractive. This occurs when
the second term in the numerator, $U(2x-1)$, which is negative for $x<0.5$,
dominates over the first one. We suggest that this second term, which 
originates from the two diagrams shown in Fig. \ref{fig:diag} {\it (c)}, is
due to the many-body effects. 

Indeed, the large-$U$ partially-filled Hubbard conduction band has (at least 
within the mean-field picture -- see the discussion in Sect. 
\ref{sec:spinwave} 
above) the ferromagnetic tendencies of its own, which are unrelated to the 
ionic spins and double exchange. In the $J_H=0$ case, the ferromagnetic, fully 
spin-polarised state of the carriers (uncoupled to the ionic spins) 
corresponds to the largest bandwidth and hence to the lowest kinetic energy. 
Let us now assume that the value of $J_H$ is finite and the ionic spins are 
fixed in a certain configuration which is not perfectly ferromagnetic 
(in the present context, the deviation of 
the ionic spin configuration from the ferromagnetic ground state 
is due to the thermal fluctuations).
The electron bandstructure is then determined by a competition between
the Hubbard band ferromagnetism (which favours a uniform ferromagnetic 
alignment of the carrier spins and hence decoupling from the ionic spin 
background) and the
Hund's rule coupling (which tends to align the carrier spins locally with the 
ionic ones, leading to the double exchange band narrowing)\cite{manybodystiff}.
 Not surprisingly for an interacting 
many-body system, the resulting 
bandstructure (along with the local carrier spin direction and the 
overall strength of band ferromagnetism) therefore depends on the bandfilling 
$x$.
As a result of electron-electron interaction, 
the mean-field bandwidth of spin-polarised carriers (somewhat suppressed 
due to the double exchange mechanism) then increases whenever the bandfilling 
of electrons (or holes) is 
increased, reaching the maximum at $x=0.5$ (where the effects of Hubbard band 
ferromagnetism are most pronounced).
The kinetic energy of an electron at a sufficiently
small value of $x$ may thus
be lowered if another electron is present nearby (a local increase of the 
carrier density), leading to the effective attraction. 

This effect is likely to have important
physical consequences in the case of Eu-based magnetic semiconductors, as
the effective attraction will further improve the stability of microscopic 
droplet-like areas with increased carrier density and enhanced ferromagnetic 
order (ferrons\cite{Nagaevbook}), which were 
argued\cite{Nagaevbook,Samokhvalov} to play a key role in 
these compounds. If the bare nearest-neighbour Coulomb repulsion is not too 
strong (which is expected), taking the effective interaction into account may
in fact
lead to the total nearest neighbour interaction being attractive 
above a certain temperature.
This in turn might signal an instability of the homogeneous ferromagnetic
state and possibly the formation of ferrons. This question obviously calls for 
further investigation.

The second term in Eq. (\ref{eq:Vreal}) is a variant of {\it correlated
electron hopping}. While the effects of this term in the present context are 
not immediately clear, we note that correlated hopping represents a 
much-studied
extension of the Hubbard model. In fact, an effective interaction somewhat
similar to (\ref{eq:Vreal}) (although involving fermions with antiparallel 
spins)  has been obtained in the 
past\cite{Hirsch} within a mean-field approach to a large-$U$ Hubbard model.
In the case of CMR manganates, where the quantity $V(T)$ is very small, the
effects of the correlated hopping term in Eq. (\ref{eq:Vreal}) are expected
to be negligible. Again, the situation can be very different for the CMR 
magnetic semiconductors with small carrier densities. 

We note that while
the result (\ref{eq:Vreal}) applies only for  degenerate magnetic
semiconductors with finite carrier densities at $T \rightarrow 0$, we
expect a similar magnon-mediated effective electron-electron interaction 
to arise in the non-degenerate (undoped) case as well.

\section{PHASE SEPARATION AND PHASE DIAGRAM AT T=0}
\label{sec:phase}

Although spin stiffness $D(x)$ and spin-wave energy, $\omega_{\vec{p}}$,
discussed in Sect. \ref{sec:spinwave}, are important and much-studied 
quantities characterising magnetic properties of double exchange ferromagnets,
their behaviour is not expected to yield any conclusive information on the 
stability of the ferromagnetic state at low temperatures. Indeed, for 
large $J_H$ and $U=0$ it is 
possible to verify (see, {\it e.g.}, Ref. \cite{jap02} and references therein)
that, in a marked difference from
conventional isotropic insulating magnets with competing interactions, 
the zero-temperature magnon  
spectrum of a double exchange magnet in the 
homogeneous ferromagnetic state does not soften when the latter is 
rendered unstable due to a change in the balance between ferro- and 
antiferromagnetic tendencies in the system. This is because phase separation,
which is a generic phenomenon found in both experimental and theoretical
studies of the CMR manganates and doped magnetic 
semiconductors\cite{Nagaevbook,Dagottobook} always preempts a second-order,
spin wave-mediated phase transition.
On the theory side, there is little doubt that this situation persists in
the finite-$J_H$, $U>0$ case in two and three dimensions (in particular,
this is known to be true in the $J_H \rightarrow 0$, large-$U$ case
of the pure Hubbard model\cite{Visscher}). In the present 
section, we will see how the presence of on-site Coulomb repulsion $U$ 
affects the physics of phase separation and, in particular, the 
low-temperature phase diagram.  We note that the effects of Coulomb repulsion
on phase separation in this  model were studied in 
Ref. \cite{Mishra} within a somewhat different mean-field approach.

At a given value of electron band filling $x$ [and the corresponding chemical
potential, $\mu(x)$], the homogeneous ferromagnetic phase of a double exchange 
magnet is unstable with respect
to phase separation whenever its thermodynamic potential,
\begin{equation}
\Omega_{FM} [\mu(x), J] = E-\mu x+ d J, \,\,\,E=\int \epsilon_{\vec{k}} 
n_{\vec{k}} \frac{d^dk}{(2 \pi)^d}\,,
\label{eq:omegafm}
\end{equation}
is larger than the thermodynamic potential $\Omega_P$ of another homogeneous 
phase $P$, 
calculated at the same value of $\mu$. The total energy of the system can 
then be lowered if areas of this new phase are formed within the bulk of
the ferromagnet\cite{shape}. Therefore in order to find the stability 
region of the homogeneous ferromagnetic phase, one has to identify the
relevant phases and  evaluate the dependence of their respective 
thermodynamic potentials on the parameters of the Hamiltonian,  
Eq. (\ref{eq:Ham}). 

For any given values of $J_H$ and $U$, it is convenient to characterise the
stability of the homogeneous ferromagnetic state by the critical value of
direct antiferromagnetic exchange coupling, $J_{cr}(x)$, above which the 
system becomes
phase-separated. The calculation thus proceeds as follows: the equation,
\begin{equation}
\Omega_{FM}[\mu(x),J_P]=\Omega_P[\mu(x),J_P]\,,
\end{equation}
is solved for several possible phases $P$, yielding the corresponding
values of $J=J_P(x,U,J_H)$. At a fixed $x$, the value  of $J_{cr}$  is 
then given
by the lowest $J_P$. Such a procedure clearly has a variational validity as
it does not imply an existence of
a rigorous proof that $J_{cr}$ cannot be lowered 
further by broadening our selection of  phases $P$.  On the other hand, the
condition $J>J_{cr}$ is obviously {\it sufficient} for the phase separation
to occur. We shall now turn to the two possible antiferromagnetic 
phases considered in the
present paper. 
 
Based on the numerous results for the $U=0$ 
case\cite{Nagaevbook,jap02,Dagottobook}, one expects that a phase separation
into ferromagnetic and the usual {\it N\'{e}el 
antiferromagnetic (G-type antiferromagnetic) phase}, corresponding to the wave 
vector of 
$\{\pi,\pi,\pi\}$ (in 2D, $\{\pi,\pi\}$) takes place in the vicinity of
endpoints, $x=0$ and $x=1$. To leading order in $1/S$, one can use the
classical formalism \{see, {\it e.g.,} Eqs. (9-10) of Ref. \cite{prb03}\}
to re-write the Hamiltonian
in terms of the new fermions $d_{i\uparrow}$ (and $d_{i\downarrow}$), whose
spins are aligned (antialigned) with the local ionic spins $\vec{S}_i$ in the 
G-type antiferromagnet:
\begin{eqnarray}
{\cal H}_G&=&-\frac{t}{2} \sum_{\langle i,j\rangle} \left( 
d^\dagger_{i \downarrow} d_{j \uparrow} + d^\dagger_{i\uparrow} 
d_{j \downarrow} + {\rm H. c.} \right) + \nonumber \\
&&+\frac{J_H}{2} \sum_i 
\left( d^\dagger_{i \downarrow}d_{i \downarrow}-
d^\dagger_{i \uparrow}d_{i \uparrow} \right)-dJN+ \nonumber \\
&&+U\sum_i\left(d^\dagger_{i \uparrow}d_{i \uparrow} x_\downarrow+
d^\dagger_{i \downarrow}d_{i \downarrow} x_\uparrow-x_\uparrow x_\downarrow 
\right)\,. 
\label{eq:HamGAFM}
\end{eqnarray}
Here, a standard Hartree mean-field decoupling has been carried out in the
last term, with $x_\alpha \equiv \langle d^\dagger_{i \alpha}d_{i \alpha}
\rangle$ 
denoting the average
spin-up and -down fermion densities. Upon Fourier transformation and 
diagonalisation (see Appendix \ref{app:GAFM}), Eq. (\ref{eq:HamGAFM}) takes 
form
\begin{equation}
{\cal H}=\sum_{\vec{k},\alpha} \epsilon_G^{\alpha}(\vec{k}) f^\dagger_{\vec{k},
\alpha} f_{\vec{k},\alpha} -dNJ -UNx_\downarrow x_\uparrow\,.
\label{eq:HamGAFM2}
\end{equation}
Here, $f^\dagger_{\vec{k},\alpha}$ are the new fermion creation operators,
the momentum summation is performed over the full (ferromagnetic) Brillouin 
zone, and the carrier spectrum is given by
\begin{eqnarray}
\epsilon_G^{\uparrow,\downarrow}(\vec{k})&=& \delta \mp \sqrt{\frac{1}{4}
(J^{(G)}_S)^2 + (\epsilon_{\vec{k}})^2}\,,
\label{eq:specG} \\
J^{(G)}_S&=&J_H+U(x_\uparrow-x_\downarrow)\,,\,\,\,\delta=
\frac{1}{2}U(x_\uparrow+x_\downarrow)\,. 
\label{eq:StonerG}
\end{eqnarray}
These equations are valid also at $U=0$, in which case $J_S^{(G)}=J_H$ and 
$\delta =0$. For general values of $U$,
the quantity $J^{(G)}_S$ represents the Stoner-type mean field band 
splitting in the
$G$-antiferromagnetic phase, whereas $\delta$ merely renormalises the
chemical potential, $\mu - \frac{1}{2}J_H$. The closed system of mean field
equations for these parameters of the $G$-antiferromagnetic phase at a given
band filling, $x_G=x_\uparrow+x_\downarrow$, is presented and discussed
in Appendix \ref{app:GAFM}. As we are interested
only in finding the phase-separation instability of the homogeneous 
ferromagnetic state, we do not need to solve these mean field equations for
general values of $x_G$. The latter will rather be determined by $\mu$, which
in turn is related in a usual way to the band filling $x$ of the homogeneous
double exchange ferromagnet. 

When the value of $x$ is small, the chemical potential in the ferromagnetic phase
lies below the bottom of the lower band of $G$-antiferromagnet [given by
Eq. (\ref{eq:specG}) with $x_\uparrow=x_\downarrow=0$],
\begin{equation}
\mu(x) < \mu_0=\frac{1}{2}J_H-\sqrt{\frac{1}{4}J_H^2+t^2 d^2}\,,
\label{eq:xGzero}
\end{equation}
and the thermodynamic potential of the N\'{e}el phase, $\Omega_G$, equals 
$-dJ$. As the value of $x$ increases, the inequality (\ref{eq:xGzero})
is eventually violated, giving rise to a non-zero carrier density $x_G$ 
(with $x_\uparrow>x_\downarrow$) in the antiferromagnetic phase. 

For the realistic values of parameters in the 2D case, we find that the 
$G$-antiferromagnetic
phase with partially-filled band, $0>x_G>1$, is not relevant in the
context of phase 
separation. This is because within the corresponding range of values of
$x$ [and $\mu(x)$], the critical value of superexchange coupling, $J_G(x)$, 
is larger than the one which corresponds to the second-order spin-wave
transition,  $J_{sw}(x)=D_0(x)S$ (where $D_0$ is the spin stiffness evaluated
at $J=0$). The peculiar properties of the 
partially-filled case, as discussed in Appendix \ref{app:GAFM}, are interesting
on their own, and may also prove relevant  in another context. Here, 
we will turn to the case when the chemical potential $\mu(x)$ lies above
the top of the filled $G$-antiferromagnetic band,
\begin {equation}
\mu(x) >
\frac{t^2 dU}{2(J_H+U)^2}\,,
\label{eq:Gfilled}
\end{equation}
resulting in $x_G=1$. Given the realistic (large) magnitude of $U$,
 the value of $J_S^{(G)} \approx J_H+U$
is then also large, and the mean field equations are readily solved with the 
help
of the $t/J_S^{(G)}$-expansion [which was used also in writing 
Eq. (\ref{eq:Gfilled})]. We find an expression
\begin{equation}
\Omega_G=-\frac{t^2 d}{2(J_H+U)}-\mu(x)-dJ\,,
\label{eq:OmegaxG1}
\end{equation}
which can be used to determine the critical value of superexchange $J_G(x)$ 
whenever the phase separation into the partially filled ferromagnetic and 
filled ($x_G=1$)   
N\'{e}el states becomes possible. As we will see below, this happens only
well above the quarter-filling of the ferromagnetic band, $x>0.5$, where
the inequality (\ref{eq:Gfilled}) is clearly satisfied.

Another phase which seems to be ubiquitous in the 2D case is the {\it A-type
antiferromagnetic} one, characterised by the wave vector $\{\pi,0\}$.
The mean field theory of the $A$-antiferromagnetic phase is formulated along
the same lines as for the N\'{e}el antiferromagnet above, where 
Eq.(\ref{eq:specG}) is now replaced with
\begin{equation}
\epsilon_A^{\uparrow,\downarrow}(\vec{k})= \delta - t\cos k_x \mp 
\sqrt{\frac{1}{4} (J^{(A)}_S)^2 + t^2 \cos^2 k_y}\,,
\label{eq:specA}
\end{equation}
and the (negative) superexchange term in the Hamiltonian is cancelled.
In spite of the higher superexchange energy, the $A$-type phase becomes more 
profitable than the N\'{e}el one because it allows for a larger gain in
the kinetic energy of carriers. Thus, it is clear that the $A$-phase becomes
relevant only when the carrier density in the $A$-type ferromagnet 
$x_A[\mu(x)]$
~(and similarly, the hole density, $1-x_A$) is not too small. In case
of large $U$, this means that the value of Stoner bandsplitting, $J^{(A)}_S$,
is much larger than the hopping matrix element, $t$.

Therefore while for  moderate values of $U$ the full system of mean field
equations for $J^{(A)}_S$ and $x_A$ must be solved, and then the thermodynamic
potential,
\begin{eqnarray}
\Omega_A=&&\int \left\{\epsilon^\uparrow_A({\vec{k}}) + 
\frac{1}{2} J_H - \mu(x) \right\} n^{(A)}_{\vec{k}}\frac{d^2k}{4 \pi^2}
- \frac{1}{4} U x^2_A + \nonumber \\
&&+\frac{1}{4U} \left(J^{(A)}_S-J_H \right)^2
\label{eq:OmA0}
\end{eqnarray}
[where $n^{(A)}_{\vec{k}}$ is the Fermi distribution function, corresponding
to the dispersion law (\ref{eq:specA}) and chemical potential $\mu(x)- J_H/2$],
must be evaluated numerically, in the large-$U$ case it is sufficient to retain
the leading-order terms in $t/J^{(A)}_S$. In this way we obtain for $U \gg t$, 
\begin{eqnarray}
J_S^{(A)} &\approx & J_H+Ux_A \approx J_H+ \frac{U}{\pi} \left[ \pi - 
{\rm arc cos} \frac{\mu(x)}{t} \right] \,,
\label{eq:JAS}\\
\Omega_A & \approx & -\frac{1}{\pi}\sqrt{t^2-\mu^2} -\frac{1}{\pi}
\left(\mu+\frac{t^2}{2J^{(A)}_S} \right) \left(\pi - 
{\rm arc cos} \frac{\mu}{t} \right)\,.
\label{eq:OmA}
\end{eqnarray}
Eqs. (\ref{eq:JAS}--\ref{eq:OmA}) hold for $|\mu(x)|<t$, which at large $U$
corresponds to partial filling of the $A$-antiferromagnetic band.

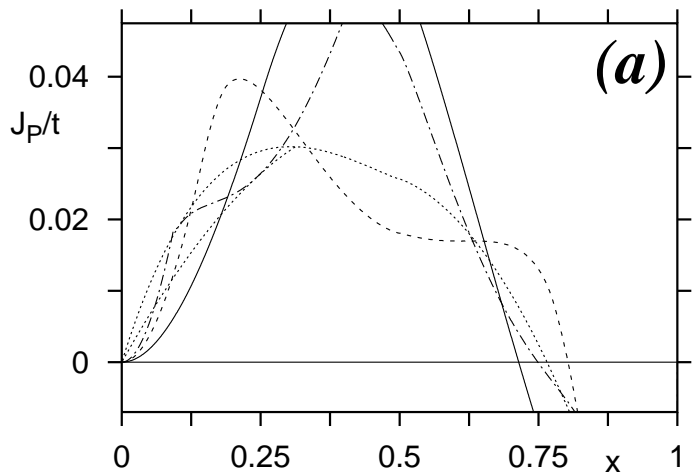
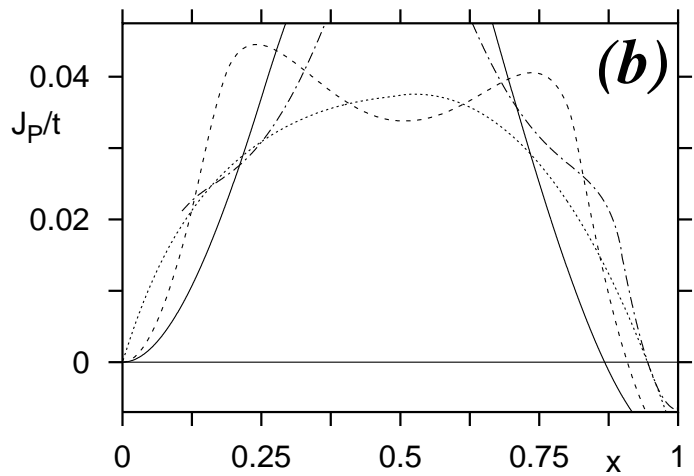
\begin{figure}
\caption{Values of superexchange $J$ corresponding to the instabilities of the
ferromagnetic order at $T=0$ in a 2D system with $J_H/t=5$, $U=0$ {\it (a)}, 
and $U/t=16$ {\it (b)}. Dotted lines correspond to spin-wave instabilities,
solid (dashed) lines -- to the phase separation into $G$-type ($A$-type) 
antiferromagnetic and ferromagnetic phases, and the dashed-dotted line -- to 
the phase separation into chain (see Fig. \ref{fig:chain}) and ferromagnetic 
phases.}  
\label{fig:phase}
\end{figure}

The zero-temperature phase diagram for a 2D system with $J_H/t=5$ and $U=0$ 
is shown in Fig. 
\ref{fig:phase} {\it (a)}. Here, the (upper) dotted line corresponds to the
long-wavelength spin-wave softening, $J_{sw}(x)=D_0(x)S$; at $x<0.31$, it
is preempted by spin wave instability at $\vec{p}=\{\pi,\pi\}$ (lower dotted
line). Phase separation instabilities involving N\'{e}el and $A$-type 
antiferromagnetic phases are represented by solid and dashed lines, 
respectively. We see that the stability region of the uniform ferromagnetic
phase is heavily shifted towards the electron-doped end, $x<0.5$; at the 
experimentally relevant values of $J/t\sim0.015$ ($J/t \sim 0.02$), the 
homogeneous ferromagnetic state is unstable everywhere at $1-x < 0.34$ 
($ 1-x <0.55$). As explained in the Introduction, it is  at
$x>0.5$ that the broad low-temperature ferromagnetic region is found for
the CMR manganates, and such an instability clearly contradicts this
experimental observation.

The situation changes in the presence of $U=16t$, as shown in Fig. 
\ref{fig:phase} {\it (b)}. Here, the dotted line represents our result for
spin stiffness, Eq. (\ref{eq:stiff}); due to the 
zone-boundary softening effect
discussed in Sect. \ref{sec:spinwave}, it is expected that for all carrier
concentrations, the spin-wave
instability at $\vec{p}=\{\pi,\pi\}$ corresponds to a slightly lower value
of superexchange. The latter is not shown in  Fig. 
\ref{fig:phase} {\it (b)} because our result for the magnon spectrum, Eq. 
(\ref{eq:magnonspec}), does not include the subleading (in $t/J_S$) term and
therefore should not be compared with other quantities plotted here.
The N\'{e}el phase separation instability (solid line) is plotted
using Eqs. (\ref{eq:xGzero}), (\ref{eq:OmegaxGsmall}), and (\ref{eq:OmegaxG1}).
In order to improve accuracy for $\mu(x)<-1$ (corresponding to $x<0.19$), 
we solved the full system of mean field equations for the $A$-type phase.
Nevertheless, we note that Eq. (\ref{eq:OmA}) yields accurate results in
the region where the phase separation into ferromagnetic and $A$-type 
antiferromagnetic phases is possible. 

We see that for $U/t=16$ and $J/t=0.015$ ($J/t = 0.02$), the region where
ferromagnetic phase is unstable is shifted to $1-x < 0.21$ ($1-x < 0.23$),
so that a {\it substantial stability region is now left for the ferromagnetic
phase at} $x>0.5$, {\it in agreement with experimental results}. 
Admittedly, the
presence of broad stability region of ferromagnetic phase at $x<0.5$ is at
variance with experiments and indicates a deficiency of either our variational
procedure ({\it i.e.}, our choice of possible phases is too narrow), or our
simplified model, Eq. (\ref{eq:Ham})\cite{eldoped}. 
Indeed, in the 3D case the experimental data\cite{Martin,Chmaissem} show
that the phase diagrams for different perovskite compounds differ in the
electron-doped ($x<0.5$) region, suggesting the sensitivity to details
of crystalline surrounding and perhaps the importance of orbital structure.

The inclusion of Hubbard interaction in principle could have brought about
new charge-ordered antiferromagnetic (or ferrimagnetic) phases, which would 
not occur at $U=0$. While we tried to look into this possibility, we could
not identify any such phases that would be stable within the experimentally
relevant region of parameter values. This difficulty was encountered also
by other workers in the field\cite{Mishra}, who could stabilise charge ordering
only upon including a large intersite Coulomb repulsion. 

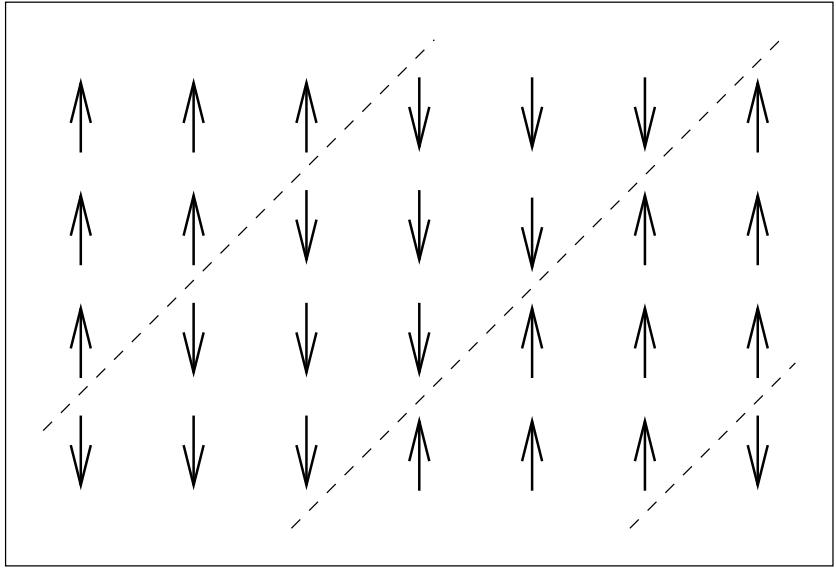
\begin{figure}
\caption{Spin ordering in the chain phase.}  
\label{fig:chain}
\end{figure}

It is however worth mentioning that some of the more complicated phases which
are relevant for phase separation at $U=0$ do show charge ordering owing
to inequivalence of different lattice sites. This is exemplified by the
{\it chain phase}\cite{jap02}, shown schematically in Fig. \ref{fig:chain}.
In Fig. \ref{fig:phase}, the critical value of superexchange, corresponding
to phase separation into ferromagnetic and chain phases, is shown by 
the dashed-dotted line; for the $U=0$ case it was calculated exactly, whereas
for $U=16t$ we used the large Stoner bandsplitting expansion similar to the
one described above for the $A$-antiferromagnetic phase. We see that when $U$
is included, the phase separation into ferromagnetic and chain phases in
the hole-doped region, $x>0.5$, becomes impossible. This is in line with the
general expectation that {\it Hubbard repulsion disfavours charge ordering
at $x>0.5$}, when there is less space between electrons. 

The present mean-field treatment allowed us to arrive at important 
conclusions regarding the effects of Hubbard repulsion on the possible 
instabilities of the homogeneous 
ferromagnetic phase. Nevertheless, we stress that  the full zero-temperature 
phase diagrams for the model (\ref{eq:Ham}) both in 2D and in a much
more cumbersome 3D case are still lacking, and should only come
from numerical experiments. This is a challenging problem, as 
some of the phases involved can be expected to have relatively large
unit cells.

\section{CARRIER DENSITY OF STATES NEAR THE FERMI LEVEL}
\label{sec:dos}

While the physical nature of carrier transport and magnetotransport in
the CMR compounds near the Curie temperature remains largely unknown, it may be
possible to single out an equilibrium property which is most closely
related to the CMR phenomenon. It appears that such a property is
a broad {\it depletion of carrier density of states near the Fermi level} 
as observed in photoemission/absorption in the CMR manganates\cite{Dessau}. 
This decrease of the density of states,
visible already deep in the ferromagnetic phase, becomes progressively
more pronounced as the temperature approaches $T_C$, around which there is
no spectral weight left at the Fermi level within the accuracy of the
experiments. These results were subsequently confirmed by the tunnelling
measurements\cite{Biswas}; it was suggested that the hard gap opening
at $T \approx T_C$ is responsible for the peak of the resistivity. The 
relevance of these gap or pseudogap phenomena for transport is underlined by
the fact\cite{Biswas} that the ``transport gap'' seen in the activated 
temperature dependence
of resistivity at $T>T_C$ is roughly of the same order of magnitude
(tenths of eV) as the width of the density of states depletion.
Furthermore, it seems possible that in the case of the non-manganate lightly 
doped or
non-degenerate magnetic semiconductors, the well-known giant red
shift of the optical absorption edge \cite{Wachter} (with optical gap 
decreasing as the temperature is lowered through $T_C$; see the 
discussion in Ref.
\cite{Nagaevbook}) may act as a counterpart
of the temperature-dependent pseudogap observed in the manganates. 

For the case of the CMR manganates, high-resolution tunnelling measurements
have recently been extended\cite{Mitra} down to liquid helium temperatures, 
revealing
a noticeable, albeit narrow, depletion of the density of states near the 
Fermi level at $T=4.2 K$. Although the conclusive evidence is still lacking,
it is most reasonable to expect that it is this feature which with increasing 
temperature evolves into the broad pseudogap observed near $T=T_C$. It is
therefore important that a proper description of the low-temperature
ferromagnetic state of the CMR compounds should include this anomaly.

Within the mean-field picture advanced in the present paper, at low 
temperatures and within the relevant doping range of $1-x \approx 0.3$,
the system is assumed to be in a homogeneous, half-metallic 
ferromagnetic state; deviations from ferromagnetic ordering (spin waves) 
freeze out at $T \ll T_C$. The resistivity of the sample is then
determined by impurity scattering, and it is only a combination of the latter
with the usual Coulomb repulsion (including the long-range component),
which could result in any density of states feature near the Fermi level.
Indeed, it is known from the work of Altshuler and Aronov \cite{AA,AAreview}
that electron-electron interactions in a diffusive conductor
generate an anomaly in the tunnelling density of states,
centered on the Fermi energy. We note that since the magnitude of effective
interaction $V_{eff}(T)$ in the manganates is much smaller than
the corresponding (nearest-neighbour) term $V_{nn}$ in the Coulomb 
repulsion (see
Sect. \ref{sec:scat}),
the effects of $V_{eff}(T)$ on the density of states can be omitted altogether.

We use the standard expression for the change in the density
of states near the Fermi level\cite{AA,AAreview}, 
\begin{equation}
\delta \nu = \frac{1}{\sqrt{2} \pi^2} \cdot \frac{|\epsilon -\mu|^{1/2}}
{(\hbar D_e)^{3/2}}\,,
\label{eq:AA}
\end{equation}
where  the overall pre-factor has been multiplied by two in order 
to account for 
half-metallicity of the system\cite{factor2}. Assuming $t\sim 0.5$eV and 
using the values $\rho \sim 161 \mu\Omega$cm for the resistivity 
of  ${\rm La_{0.7} 
Ca_{0.3}Mn O_3}$ \cite{Mitra} and $a\sim 4 {\rm \AA}$ for intersite distance,
we estimate the diffusion
constant $D_e$  as $D_e\sim [\rho e^2 \nu(\epsilon_F)]^{-1} 
\sim 6 a^3 t/(\rho e^2)\sim 7 {\rm cm^2/s}$. 
The resulting estimate, 
\begin{equation}
\frac{\delta \nu(\epsilon)}{\nu} \sim 0.03 \sqrt{\frac{|\epsilon - \mu|}{t}}
\label{eq:dosestimate}
\end{equation}
is an order of magnitude smaller than the experimental results of 
Ref. \cite{Mitra}, which show a 15\% change in $\delta \nu/\nu$ for 
$|\epsilon - \mu|\sim 0.075$eV. Furthermore, based on Eq. (\ref{eq:AA}) 
one  expects that the relative change in the density of states for
$\rm{La_{0.75} Sr_{0.25} Mn O_3}$ (characterised by smaller values of
resistivity and by a larger bandwidth) should be about 10 times less than in 
the case of  ${\rm La_{0.7} Ca_{0.3}Mn O_3}$, whereas experimentally 
the two curves differ by a factor of the order of 2. In addition, the 
experiments\cite{Mitra} yield $\delta \nu/\nu \propto (\epsilon-\mu)^2$ 
within a relatively broad energy range of
$|\epsilon-\mu|<0.02$eV; this is in contrast with the standard 
theory\cite{AA,AAreview}, which predicts a crossover to the square root
law [cf. Eq. (\ref{eq:AA})] at $|\epsilon-\mu| \sim T$. 
Thus, we find that {\it the Altshuler -- Aronov
mechanism cannot possibly account for the Fermi-level density of states 
depletion in the CMR manganates} even at low temperatures.

We therefore conclude that the low-temperature ferromagnetic state of
the CMR manganates is characterised by strong electron correlation 
effects. Since these are certainly not captured within the present
mean-field approach this does not necessarily signify the deficiency
of our simplified model, Eq. (\ref{eq:Ham}). While we plan to investigate
this question in more detail in the future, we emphasise that our
conclusion on the {\it correlated nature of the low-temperature 
ferromagnetic phase} of the CMR manganates is likely to be model-independent.
In other words, an adequate generic model of the manganates, whether or not 
it involves orbital, lattice, {\it etc}. degrees of freedom, must necessarily
take proper non-perturbative account of electron-electron interactions.

\section{CONCLUSION}
\label{sec:conclu}

In the present paper, we were concerned with the effects of the strong
on-site Coulomb repulsion which is present in the CMR compounds but often
overlooked in the theoretical treatments. By treating the model within the
mean field approach, we were able to resolve some apparent discrepancies 
between the generically observed low-temperature properties of these compounds
and theoretical results for the (non-interacting) double exchange model at 
the appropriate values of Hund's rule coupling strength. These properties
include the doping dependence of spin stiffness, and the ``zone-boundary
softening'' of magnon spectrum (Sect. \ref{sec:spinwave}) which has attracted
much attention from both theorists and experimentalists. The underlying 
physical mechanisms are two-fold and include both the interaction-induced
increase in the effective band splitting (Sect. \ref{sec:intro}) and
the correlated physics of the strongly interacting Hubbard carriers (Sect. 
\ref{sec:scat}). In addition, we showed that a novel, magnon-mediated
effective electron-electron interaction arises in these systems at finite
temperatures (Sect. \ref{sec:scat}). While for  the CMR 
manganates the strength of this interaction remains negligible, it is
expected that it is much more important in the case of Eu-based magnetic
semiconductors exhibiting CMR. 

By considering the stability of the ferromagnetic state against phase separation,
we were able to show (Sect. \ref{sec:phase}) that inclusion of the Hubbard 
repulsion alleviates
another disagreement between the theory and experiment, resulting in a
sizable stability region of the ferromagnetic 
state above half electron filling, $x>0.5$.  
Regarding the phase diagram, the question of identification of the 
relevant phases and finding the precise domain of the ferromagnetic 
phase (especially in the electron-doped region, $x < 0.5$)
remains open and calls for further theoretical investigations, in particular
numerical ones. At the same time we note that the underlying physics
consists in a competition between many phases with very close values
of thermodynamic potential, and the outcome is guaranteed to be strongly
dependent on the details of band structure, lattice/orbital properties
and interactions in a particular compound. Therefore, while understanding
the details and implications of phase separation in the CMR compounds 
(including both thermodynamic and transport properties) presents a broad 
and fascinating
problem, it is not obvious that these details are directly related to the 
generic features of the CMR phenomenon itself. 

The satisfactory results of our mean field approach, as sketched above, all
have to do with the integral quantities, involving summation over the entire
Fermi sea. It is precisely this effective averaging that makes our 
Hartree--Fock decoupling scheme a relatively reliable tool in this case.
The situation changes drastically when this approach is used to address
other issues, such as the behaviour of the carrier density of states
near the Fermi level (Sect. \ref{sec:dos}). In this case, the usage of the
Hartree--Fock approximation (which yields the effective Stoner band splitting
$J_S$ much larger than the Fermi energy, leading to an assumption that
the system is half-metallic with no minority-spin carriers) is the probable
cause of our inability to reproduce the experimentally measured\cite{Mitra} 
depletion
of the density of states. Assuming that the results reported  in 
Ref. \cite{Mitra} are sufficiently generic, this failure may have 
far-reaching conceptual consequences.

As mentioned in Sect. \ref{sec:dos}, it is expected that the low-temperature
depletion of the density of states at the Fermi level is a 
temperature-dependent feature, which with increasing temperature evolves into
the pseudogap; this temperature dependence is in turn  expected to crucially 
affect transport properties of
the system both near $T_C$ and at low temperatures. The failure to recover
the low-$T$ Fermi-level feature in the density of states within a theoretical 
treatment 
based on the picture of a half-metallic homogeneous ferromagnetic phase
(Sect. \ref{sec:dos}) should lead to {\it questioning the experimental relevance
of the many available calculations of the low-temperature resistivity in
double exchange ferromagnets}, which are based on similar assumptions.

On the other hand, one should not overlook the evidence, coming both from
band structure calculations\cite{Pickett} and the experimental 
observations\cite{Nadgorny}, which points to the {\it presence of carriers in 
the minority spin subband} in the CMR manganates even at low temperatures. 
Theoretically, this may 
be possible due to relatively small values of Hund's rule coupling; the 
spin-down
(minority) electrons\cite{t2g}, if present, will form  polaron-like localised 
states, 
accompanied by a reduction in the spin-up electron density within the area of
the polaron (this reduction will in turn reduce the Coulomb/Hubbard 
interaction energy). 
These localised spin-down carriers will clearly lead to an 
enhanced spin-up electron scattering and will also affect the density of states
at the Fermi level\cite{polarons}. As explained 
in the Introduction, these correlated effects are beyond the mean-field 
approach used in the present paper; in our view, this scenario definitely
merits further attention, 
especially as there is now some recognition\cite{Michaelis} that 
other avenues of theoretical investigation of the CMR and related phenomena 
may have proved unpromising. 

On the experimental side, we suggest that some key measurements still
have to be performed in order to clarify the issues under discussion here.
These fall into three categories:

\noindent (i) Detailed investigations of the energy dependence of the density of states
as a function of temperature and magnetic field. This includes tunnelling 
and optical measurements for various chemical composition and doping levels,
within the entire temperature range, and would help to clarify the relationship
between the low-temperature density of state depletion of the Fermi 
level\cite{Mitra} and the pseudogap observed in the near-critical 
region\cite{Dessau,Biswas}, as 
well as confirm the relevance of these phenomena for transport and 
magnetotransport. 

\noindent (ii) Systematic investigation of the interplay between pseudogap 
and other phenomena, in particular those 
related to the unusual spin correlations found in the CMR manganates, such as 
the central peak observed 
in the inelastic neutron scattering\cite{Lynn} and critical behaviour of 
spin-stiffness\cite{Ferbaca}.
In particular, it can be expected that understanding the nature of spin 
dynamics at elevated temperatures would shed light on the structure of
electron states involved in the pseudogap formation. The relationship
between pseudogap and phase separation should be clarified as well.  

\noindent (iii) Lastly, what appears an important theoretical  and 
experimental problem is
to  identify the ``common denominator'' between the structure
and properties of the CMR manganates\cite{Tokurabook} and those of the
CMR magnetic semiconductors (Eu-based) and spinels\cite{Nagaevbook}. 
Understanding what  these CMR
compounds have in common may be of much help in constructing a minimal
theoretical model for the CMR compounds.
In connexion to this, we recall that the familiar argument concerning the
relative unimportance of $U$ in  systems with low carrier densities rests 
upon an
assumption that carrier distribution throughout the sample is uniform on 
the microscopic scale.
The latter is not expected to be true, both for the CMR manganates and for
ferromagnetic semiconductors at elevated temperatures, 
$T\stackrel{<}{\sim} T_C$. It is therefore possible that the effects
of $U$, discussed in this paper, are important in both systems.

\acknowledgements

I am indebted to J. T. Chalker for the many detailed discussions which provided
me with his insight and counsel throughout all stages of work on
the present article. I also take pleasure in thanking A. M. Finkelstein,
K. A. Kikoin, J. W. Lynn, and S. Satpathy for valuable discussions. 
This work was  supported by
the EPSRC Grant No. GR/M0442, by the ISF of the Israeli Academy, 
by the EC RTN Spintronics, and by the Israeli Ministry of Absorption.

\appendix
\section{Mean Field Equations for the G-type Antiferromagnetic Phase at T=0}
\label{app:GAFM}

Here we outline the necessary details of  the Hartree mean field scheme, 
as applied in Sect. \ref{sec:phase} to the $G$-antiferromagnetic phase.
Throughout this appendix, we use units in which
the hopping constant, $t$, is equal to unity.
Eq. (\ref{eq:HamGAFM}) is diagonalised by
\begin{eqnarray}
d_{\vec{k} \uparrow} &=& \frac{\epsilon_{\vec{k}}}
{\sqrt{2\Omega_{\vec{k}}^2-J^{(G)}_S \Omega_{\vec{k}}}}f_{\vec{k}\uparrow}+
\frac{\epsilon_{\vec{k}}}
{\sqrt{2\Omega_{\vec{k}}^2+J^{(G)}_S \Omega_{\vec{k}}}}f_{\vec{k}\downarrow},
 \nonumber \\
d_{\vec{k} \downarrow} &=& \frac{\frac{1}{2}J^{(G)}_S-\Omega_{\vec{k}}}
{\sqrt{2\Omega_{\vec{k}}^2-J^{(G)}_S \Omega_{\vec{k}}}}f_{\vec{k}\uparrow}+
\frac{\frac{1}{2}J^{(G)}_S+\Omega_{\vec{k}}}
{\sqrt{2\Omega_{\vec{k}}^2+J^{(G)}_S \Omega_{\vec{k}}}}f_{\vec{k}\downarrow} 
\end{eqnarray}
with $\Omega_{\vec{k}}^2=\frac{1}{4}(J^{(G)}_S)^2+\epsilon^2_{\vec{k}}$.
Assuming that the system is still half-metallic, {\it i.e.} that the
chemical potential lies below the bottom of the spin-down 
{\it antiferromagnetic} 
band [cf. Eq. (\ref{eq:specG})], we 
find
\begin{equation}
x_{\uparrow, \downarrow}=\frac{1}{2}x_G\pm \frac{1}{2}I\,,\,\,\,I =\frac{1}{2N}
J^{(G)}_S\sum_{\vec{k}} \frac{n^{(G)}_{\vec{k}}}{\Omega_{\vec{k}}}\,,
\label{eq:GAFMmean}
\end{equation}
where $n^{(G)}_{\vec{k}}\equiv
\langle f^\dagger_{\vec{k}\uparrow} f_{\vec{k}\uparrow} \rangle$ is the 
appropriate Fermi distribution function. Together with Eqs.(\ref{eq:specG}--
\ref{eq:StonerG}), Eq.(\ref{eq:GAFMmean}) forms a closed system of mean
field equations for a homogeneous $G$-antiferromagnetic phase at fixed $x_G$.

We note that the Fermi surface in a partially-filled spin-up band has two 
sheets, corresponding to different signs of $\epsilon_{\vec{k}}$. Thus,
the quantity
\[ \langle d^\dagger_{i\uparrow} d_{i \downarrow} \rangle=
-\frac{1}{2N}\sum_{\vec{k}}\frac{\epsilon_{\vec{k}} n^{(G)}_{\vec{k}}} 
{\Omega_{\vec{k}}} \]
is always equal to zero, showing the consistency of mean field decoupling
used in Eq. (\ref{eq:HamGAFM}).

When the carrier density in the ferromagnetic phase becomes sufficiently 
large for the inequality (\ref{eq:xGzero}) to be  violated, there arises
a non-zero carrier density, $x_G=x_\downarrow+ x_\uparrow$ in the N\'{e}el
phase. 
This in turn
leads to an increase of quantities $\delta$ and $J^{(G)}_S$ in Eq. 
(\ref{eq:specG}), resulting in the upward shift and narrowing of the
antiferromagnetic band. This effect is more pronounced when the value of $U$
is sufficiently large, in which case the energy difference between
chemical potential and the bottom of $G$-antiferromagnetic band,
\begin{equation}
\zeta=\mu(x) -\frac{1}{2}J_H-\delta+\sqrt{\frac{1}{4}(J_S^{(G)})^2+d^2}\,,
\label{eq:defzeta}
\end{equation}
and the band-filling,
\begin{equation}
x_G=\left\{ \begin{array}{ll}
\frac{(2 \zeta)^{3/2}}{3^{5/2}\pi^2} \left[\frac{1}{4}\left(J_S^{(G)}\right)^2
+9\right]^{3/4} & 
\mbox{in 3D,}\\ & \\ \frac{\zeta}{2\pi} \sqrt{\frac{1}{4}
\left(J_S^{(G)}\right)^2
+4}& \mbox {in 2D,}
\end{array} \right.
\label{eq:xGzeta}
\end{equation}
remain small within an extended range of values of $x$. In this large-$U$ 
limit, it is possible to solve the mean-field equations analytically.

When the band filling is small, $x_G \ll 1$, Eqs. (\ref{eq:GAFMmean}) and
(\ref{eq:StonerG}) yield
\begin{equation}
J_S^{(G)}=J_H+UJ_S^{(G)}\frac{x_G}{\sqrt{(J_S^{(G)})^2+4d^2}}\,,
\end{equation}
or 
\begin{equation}
J_S^{(G)}=\frac{J_H}{1-U\xi/2}\,,\,\,\,\,\,\,\,\xi\equiv \frac{2x_G}
{\sqrt{(J_S^{(G)})^2+4d^2}}\,.
\label{eq:defxi}
\end{equation}
Substituting this into Eq. (\ref{eq:defzeta}) [and also using 
Eq. (\ref{eq:StonerG}) for $\delta$], we obtain
\begin{equation}
\zeta + \frac{1}{2}J_H-\mu(x)=\sqrt{\frac{1}{4}J_H^2+d^2\left(1-\frac{1}{2}U\xi
\right)^2 }
\label{eq:findxi}
\end{equation}
We are interested in the large-$U$ situation when $\zeta$ is small and can
be omitted on the l.\ h.\ s. [see below, Eq. (\ref{eq:accuracy})], in which 
case we find, to leading order,
\begin{equation}
\xi \approx \frac {2}{U} \left\{1-\frac{1}{d} \sqrt{\mu(x)[\mu(x)-J_H]}
\right\}.
\end{equation}
With the help of Eqs. (\ref{eq:defxi}) this in turn yields
\begin{equation}
J^{(G)}_H=dJ_H/\sqrt{\mu(x)[\mu(x)-J_H]}\,.
\label{eq:StonerGres}
\end{equation}
While at $\mu(x)=\mu_0$, Eq. (\ref{eq:StonerGres}) yields $J^{(G)}_S=J_H$,
with a further increase of $x$ [and hence $\mu(x)$] towards the 
quarter-filling,
$\mu(x)=0$, the value of $J^{(G)}_S$ increases, and the bandwidth [of the order
of $d^2/(J^{(G)}_S)^2$ , see Eq. (\ref{eq:specG})] decreases, with the effect
that the bottom of spin-up band in the $G$-antiferromagnetic phase remains
pinned immediately below the chemical potential $\mu(x)-\frac{1}{2}J_H$.
As a result, band-filling in the antiferromagnet,
\begin{equation}
x_G=\frac{2}{U} \left\{ \frac{d}{\sqrt{\mu(x)[\mu(x)-J_H]}}-1 \right\}\cdot
 \left\{\frac{1}{2}J_H-\mu(x)\right\}\,,
\label{eq:xG}
\end{equation}
remains small as long as $|\mu(x)| \gg J_H/U^2$.
 When the latter inequality 
is violated, Eqs. (\ref{eq:StonerGres}--\ref{eq:xG}) become invalid. 
Using Eqs. (\ref{eq:HamGAFM2}), (\ref{eq:specG}), and (\ref{eq:defzeta}),
we write for the thermodynamic potential in the case of small $x_G$,
\begin{eqnarray}
&&\Omega_G=-Ux_\uparrow x_\downarrow -dJ+ \frac{1}{N} \sum_{\vec{k}}
n^{(G)}_{\vec{k}}\times
\label{eq:OmG}\\
&& \times\left\{ \sqrt{\frac{1}{4}\left( J^{(G)}_S
\right)^2+d^2} -\sqrt{\frac{1}{4}\left( J^{(G)}_S\right)^2+
\epsilon^2_{\vec{k}}}- \zeta\right\} \,.  
\nonumber
\end{eqnarray}
The sum on the r.\ h.\ s. [which can be evaluated using the small-$k$
expansion and Eq. (\ref{eq:xGzeta})] is found to be of the order of
$x_G \zeta$ and can be omitted. Eqs. (\ref{eq:GAFMmean}) and (\ref{eq:xG})  
then yield the final expression,
\begin{equation}
\Omega_G=- \frac{1}{U} \left\{ d-\sqrt{\mu(x)[\mu(x)-J_H]} \right\}^2-dJ\,. 
\label{eq:OmegaxGsmall}
\end{equation}

It is easy to see that our neglecting the first term on the l.\ h.\ s. of Eq.
(\ref{eq:findxi}) is appropriate as long as
\begin{equation}
\zeta \frac{J_H-2 \mu(x)}{d\sqrt{\mu(x)[\mu(x)-J_H]}} \ll 1- \frac{1}{d}
\sqrt{\mu(x)[\mu(x)-J_H]}\,.
\label{eq:accuracy}
\end{equation}
At small absolute values of $\mu(x)<0$, this is satisfied as long as $x_G \sim
J_H^{1/2} |\mu|^{-1/2} /U$ is small [see Eqs. (\ref{eq:xG}) and 
(\ref{eq:xGzeta})]. On the other hand, when  $\mu(x)$ is close to $\mu_0$
\{when Eq. (\ref{eq:xG}) yields $x_G \approx 
[\mu(x)-\mu_0](J_H^2+4 d^2)/(Ud)$\}, 
Eq. (\ref{eq:accuracy}) takes form $U \gg \pi (J_H^2+16)^{1/2}$ in 2D and 
$U[\mu(x)-\mu_0]^{1/2} \gg \pi^2(\frac{1}{4}J_H^2+9)^{1/4}$ in 
3D\cite{validity}. 
Since the actual value of $U/t$ for the CMR manganates is about $16$, our 
results in the latter region can be viewed as an order of magnitude 
estimate only. 

As the value of $x$ is increased towards half-filling, and the chemical 
potential reaches small (negative) values of 
$\mu(x)\stackrel{>}{\sim} -J_H/U^2$, the value of $x_G$ starts increasing 
rapidly. The spin-up G-antiferromagnetic band is filled, $x_G=1$, for $\mu(x) >
dU/[2(J_H+U)^2]$ [cf. Eq. (\ref{eq:specG})]. In principle, the mean
field equations can also be analyzed within this narrow area near $\mu(x)=0$, 
assuming that 
$J_H+Ux_G$ is large.

\begin{references}
\bibitem{formal} Due to these, the spin-up band in the antiferromagnetic phases
at finite $J_H$ is not symmetric with respect to 
$\epsilon^\uparrow=-J_H/2$ [see Eq.(\ref{eq:specG}) with $\delta=0$, 
$J_S^{(G)}=J_H$]. As the spins deviate from 
ferromagnetic
ordering, the double-exchange band narrowing effect is then complemented 
by the downward shift of the centre of gravity of the band.
\bibitem{accurate} In making this generic observation, we are not concerned
with (i) the vicinity of the quarter-filling, $x=1/2$, (ii) the distinction
between metallic and insulating ferromagnetic states, and (iii) $c$-axis 
magnetic ordering in the layered CMR compounds.
\bibitem{nohole} This point of commensurate filling may prove singular from
the viewpoint of lattice and orbital effects, which are not considered here.
\bibitem{single} For an upper energy bound, consider a variational state, in 
which the spin-down particle is localised at a certain site (single-particle 
energy $J_H$), and all the spin-down electron states are perturbed in such
a way that their wave functions vanish on this site (combined single-particle
energy $\sim \epsilon_F$ with no many-body contribution).
\bibitem{highercoll} The change of ${\cal N}_{\vec{q}}$ due to the off-diagonal
terms in ${\cal H}^\prime_2$ and ${\cal H}^\prime_{i2}$ is an effect of 
higher order in $1/S$.
\bibitem{leading} As a result of our choice of $W_{\vec{k},\vec{p}}$, 
Eq. (\ref{eq:condW}), 
the leading-order (in $\epsilon_F/J_S$) contribution to magnon self-energy 
is entirely contained in ${\cal H}^\prime_2$, while other terms account
for smaller corrections. The latter, when combined with a subleading
(in $\epsilon_F/J_S$) contribution from ${\cal H}^\prime_2$, yield the last term
in Eq. (\ref{eq:stiff}).
\bibitem{exact} Even though the condition $J_S \gg \epsilon_F$ is then
violated at sufficiently large $x$, one can still use Eq. (\ref{eq:stiff}).
This is a special property of the $U=0$ case \{cf. Ref. \cite{prb03}, 
Eq. (4)\}.
\bibitem{opposite} In the opposite case of $J_H \gg Ux$, the standard 
(Heisenberg-like) expression for spin wave energy in
a double exchange ferromagnet with $J_H \rightarrow \infty$ is recovered.
\bibitem{reduced} In reality, the saturation value of $\omega_{\vec{p}}$ may
be reduced (and may also acquire some momentum dependence) if  the 
fully spin-polarised state of the Hubbard model becomes
unstable (see the preceding paragraph).
\bibitem{corr} We note that quantum-spin corrections\cite{prl00} and the
subleading terms in the $\epsilon_F/J_S$ expansion both have non-Heisenberg momentum
dependence. These effects, however, are not included in Eq. 
(\ref{eq:magnonspec}).
\bibitem{order} This contribution is of the same order in both $\epsilon_F/J_S$ and $1/S$,
but is smaller than the first one when the bare interaction $U$ is small.
\bibitem{retard} Our mean-field approach overestimates the retardation
energy scale, which in reality may be reduced from $J_S$ to
$J_H+ \epsilon_F$ (cf. footnote \cite{single}), provided that the latter is 
smaller.
\bibitem{eldoped} This difficulty was noted and discussed in
Ref. \cite{Mishra}.
\bibitem{factor2} This is because in a half-metal, all pairs of electrons
have parallel spins, thus contributing to the exchange term in the electron
self energy.
\bibitem{t2g} The minority electrons are expected to belong to the spin-down
$t_{2g}$ bands. These are likely to correspond to lower energies  
than the spin-down component of the $e_g$  band, represented
throughout the present paper by the operators $c^{\dagger}_\downarrow$.
\bibitem{polarons} We note that possible relevance of the polarons for the
density of state depletion has been discussed, although in a different context,
already in Ref. \cite{Mitra}.
\bibitem{validity} When the latter inequality is reversed in the immediate
vicinity of $\mu(x)=\mu_0$, interaction effects become unimportant and the
usual 3-dimensional behaviour of $x_G \propto [\mu(x)-\mu_0]^{3/2} $ [cf. 
Eq.(\ref{eq:xGzeta})] is restored.
\end{references}
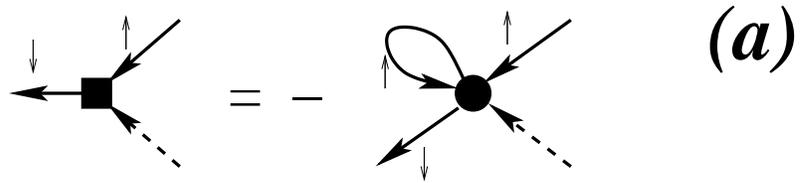
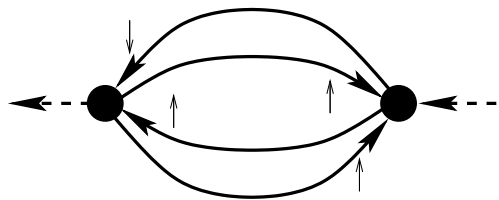
\end{document}